\documentclass[%
aip,
amsmath,amssymb,
preprint,%
]{revtex4-2}

\usepackage{graphicx}
\usepackage{dcolumn}
\usepackage{bm}
\usepackage{physics}
\usepackage{color}
\usepackage[utf8]{inputenc}
\usepackage[T1]{fontenc}
\usepackage{mathptmx}
\usepackage{setspace}
\usepackage[margin=1in]{geometry}
\usepackage{physics}
\usepackage{color}
\usepackage{multirow}
\usepackage{float}
\usepackage{adjustbox}
\usepackage{booktabs, tabularx} 
\newcolumntype{C}{>{\centering\arraybackslash}X}

\newcommand*{\vt}[1]{\textcolor{black}{ #1}}
\newcommand*{\si}[1]{\textcolor{black}{ #1}}
\newcommand*{\AST}[1]{\textcolor{black}{ #1}}

\begin{document}

\title{Rapid Scan White Light Two-dimensional Electronic Spectroscopy with 100 kHz Shot-to-Shot Detection}

\author{Asha S. Thomas}
\altaffiliation{The authors contributed equally.}
\affiliation{ 
	Solid State and Structural Chemistry Unit, Indian Institute of Science, Bangalore, Karnataka 560012, India
}%
\author{Vivek N. Bhat}%
\altaffiliation{The authors contributed equally.}
\affiliation{ 
	Solid State and Structural Chemistry Unit, Indian Institute of Science, Bangalore, Karnataka 560012, India
}%
\author{Vivek Tiwari}
\email{Author to whom correspondence should be addressed:vivektiwari@iisc.ac.in}
\affiliation{ 
	Solid State and Structural Chemistry Unit, Indian Institute of Science, Bangalore, Karnataka 560012, India
}%



\begin{abstract}
We demonstrate an approach to two-dimensional electronic spectroscopy (2DES) that combines the benefits of shot-to-shot detection at high-repetition rates with the simplicity of a broadband white light continuum input and  conventional optical elements to generate phase-locked pump pulse pairs. We demonstrate this through mutual synchronization between the laser repetition rate, acousto-optical deflector (AOD), pump delay stage and the CCD line camera, which allows rapid scanning of pump optical delay synchronously with the laser repetition rate while the delay stage is moved at a constant velocity. The resulting shot-to-shot detection scheme is repetition rate scalable and only limited by the CCD line rate and the \vt{maximum stage velocity}. Using this approach, we demonstrate measurement of an averaged 2DES absorptive spectrum in as much as \vt{1.2 seconds of continuous sample exposure per 2D spectrum}. We achieve a signal-to-noise ratio (SNR) of 6.8 for optical densities down to 0.05 with 11.6 seconds of averaging at 100 kHz laser repetition rate. Combining rapid scanning of mechanical delay lines with shot-to-shot detection as demonstrated here provides a viable alternative to acousto-optic pulse shaping (AOPS) approaches that is repetition-rate scalable, has comparable throughput and sensitivity, and minimizes sample exposure per 2D spectrum with promising micro-spectroscopy applications.
\end{abstract}

\maketitle


\section{Introduction}\label{intro}
Electronic relaxation in the condensed phase proceeds through several overlapping vibrational-electronic manifolds on femtosecond to picosecond timescales. Such phenomena span biological proteins to emerging energy materials, and carry both fundamental and applied significance. For example, sub-100 fs cis-trans photoisomerization of retinal\cite{Mathies1991} in the mammalian visual pigment rhodopsin initiates vision, and sub-50fs carrier thermalization\cite{Richter2017} may limit hot-carrier extraction in photovoltaic devices based on bulk perovskites. \\

A broadband white light continuum (WLC) light source is naturally desirable in order to probe the entire energetic manifold subsequent to a narrowband pump excitation as is typically implemented in pump-probe (PP) spectrometers \cite{Lang2018}. However, broad overlapping electronic resonances in the condensed phase along with ultrafast relaxation timescales impose the requirements of high temporal resolution with a broadband pump spectrum, and consequently also the need to know the pump excitation frequency information in order to deconvolute the underlying photophysics into a uniquely determined rate model\cite{Dostal2016}. In this regard, two-dimensional electronic spectroscopy (2DES) goes beyond conventional PP implementations in that there is no trade-off between temporal resolution and pump excitation frequency information. The spectral information is resolved in the form of a 2D contour map that correlates the initial excitation to the final detection frequency, and evolves with the pump-probe waiting time $T$. 2DES has revealed energetic relaxation pathways in complex spectrally congested systems such as protein networks within photosynthetic cells\cite{Dostal2016,Dahlberg2017}, molecular aggregate-plasmon-plexiton states\cite{Shapiro2021}, and carbon nanotube thin films\cite{Mehlenbacher2015}. Recent several fully-collinear implementations\cite{OgilvieARPC,Tiwari2021} of 2DES have also added sub-micron spatial-resolution as an additional handle to decongest the ensemble-averaged ultrafast dynamics through micro-spectroscopy. This has led to a general interest towards the development of high-repetition rate, high-throughput approaches to 2D micro-spectroscopy which also minimize sample exposure.\\

Broadband light sources in PP and 2DES have been typically\cite{OgilvieARPC} obtained through multi-stage optical parametric amplifiers (OPAs) which offer tunable\cite{Cerullo2008} few-cycle optical pulses with over 300 nm bandwidth and several hundred nanojoules (nJ) pump pulse energies \cite{Kobayashi2002,Riedle2010}. This is often complemented with a significantly simpler WLC based light source\cite{Tekavec2009,Song2019} to probe the relaxation dynamics in a broadband UV-visible-NIR region. Along similar lines, pump pulses that are also generated from a WLC are highly desirable because they can provide nearly octave spanning excitation axis\cite{Mehlenbacher2015} in 2DES without the cost and complexity of OPAs. For WLC generation through the use of nonlinear crystals\cite{Bradler2009a}, however, this also brings in significant challenges associated with power stability\cite{Bradler2009a} across the bandwidth, spectral and temporal correlations\cite{Bradler2014}, and most significantly, pulse energies of only a few tens of picojoules (pJ). Zanni and co-workers have pioneered\cite{Mehlenbacher2015,Son2022} a YAG-WLC based approach to 2DES, with extensions to high-repetition rate through shot-to-shot acousto-optic pulse shaping\cite{Kearns2017} (AOPS) at 100 kHz. YAG-WLC based 2DES approaches have also been recently implemented in action-detected variant of 2DES\cite{Kunsel2019,Ogilvie2021,Sahu2023}. Note that several 2DES approaches have been demonstrated\cite{Engel2014, Harel2015, Cohen2017} using the gas filamentation approach to continuum generation which provides $\mu$J pulse energies starting from mJ fundamental pulse energy. In comparison, a YAG-WLC only provides $\sim$1 nJ pulse energies starting with $\sim 1 \mu$J fundamental pulse energy\cite{Bradler2009a}.  \\

Shot-to-shot PP\cite{Brixner2014,Bhat2023} or 2DES\cite{Kearns2017} data collection is highly desirable because it utilizes the full laser repetition rate, leverages the shot-to-shot correlations between laser pulses to replace multichannel referencing\cite{Lang2018}, and suppresses the 1/$f$ laser noise encountered during a delay scan\cite{Moon1993}. The latter point was demonstrated in the WLC-2DES study of Kearns et al.\cite{Kearns2017} which shows signal-to-noise ratio (SNR) enhancement beyond that guaranteed by the scaling of repetition rate from 1 kHz to 100 kHz due to an additional suppression of 1/$f$ noise component. While programmable AOPS technology\cite{Tull1997,Weiner2011} has proven highly effective for high-repetition rate shot-to-shot 2D spectroscopy in the visible\cite{Kearns2017} and mid-IR\cite{Krummel2016}, significant cost and complexity, limited time aperture--RF bandwidth product, and the RF waveform update rate\cite{Tull1997} in the modulator poses limitations in terms of repetition rate scalability desirable for 2DES micro-spectroscopy\cite{Jones2020,Martin2020,Goetz2018,Tiwari2018a} applications. Development of alternate repetition rate scalable approaches to shot-to-shot WLC-2DES that rely on conventional optics \vt{and provide comparable throughput and sensitivity} is therefore quite essential in this regard.  \\

We demonstrate a repetition-rate scalable WLC-2DES spectrometer which relies on conventional optical and electronic elements to achieve repetition-rate scalable shot-to-shot data collection. The pump pulse pair is generated using birefringent-wedge based common path interferometer\cite{Cerullo2014}. 100 kHz shot-to-shot detection is achieved by mutual synchronization of the laser repetition rate, acousto-optic deflector (AOD), pump delay stage and the CCD line camera, such that the pump delay axis can be raster scanned synchronously with the laser repetition rate, while the CCD records every probe laser shot. As we have recently shown \cite{Bhat2023} in the context of WLC-PP spectroscopy, combining rapid mechanical delay scan with shot-to-shot detection provides advantages of not only increased averaging by substantially minimizing single scan time but also in suppressing\cite{Moon1993} the 1/$f$ component of experimental noise encountered during a scan. \vt{Zanni and co-workers have shown\cite{Kearns2017} that AOPS approaches to 2DES with shot-to-shot data collection can fully leverage correlations between laser shots to suppress 1/$f$ laser noise.} Our approach demonstrates that the above advantages are also possible with rapid scanning of mechanical delay lines and conventional optical elements, with an additional vital feature of repetition-rate scalability, which is in principle only limited by the camera line rate, without sacrificing pump WLC bandwidth. Overall, we demonstrate measurement of averaged 2DES absorptive spectra in as much as 1.2 seconds of continuous sample exposure per 2D spectrum, limited only by the \vt{maximum stage velocity}. We achieve an SNR of 6.8 for ODs down to 0.05 in 11.6 seconds of averaging at 100 kHz laser repetition rate, demonstrating \vt{throughput and sensitivity comparable to that reported for AOPS approaches}\cite{Kearns2017}. Overall, we introduce a considerably simpler and viable alternative to WLC-2DES that is repetition rate scalable and minimizes sample exposure per 2D spectrum with promising applications in 2DES micro-spectroscopy.

\section{Experimental Methods}

This section describes the experimental setup, interferometer for pulse pair generation and electronics synchronization scheme for shot-to-shot detection, and compares the various data acquisition and averaging schemes.

\subsection{Experimental Setup} 

\begin{figure}[h!]
	\centering\includegraphics[width=14cm]{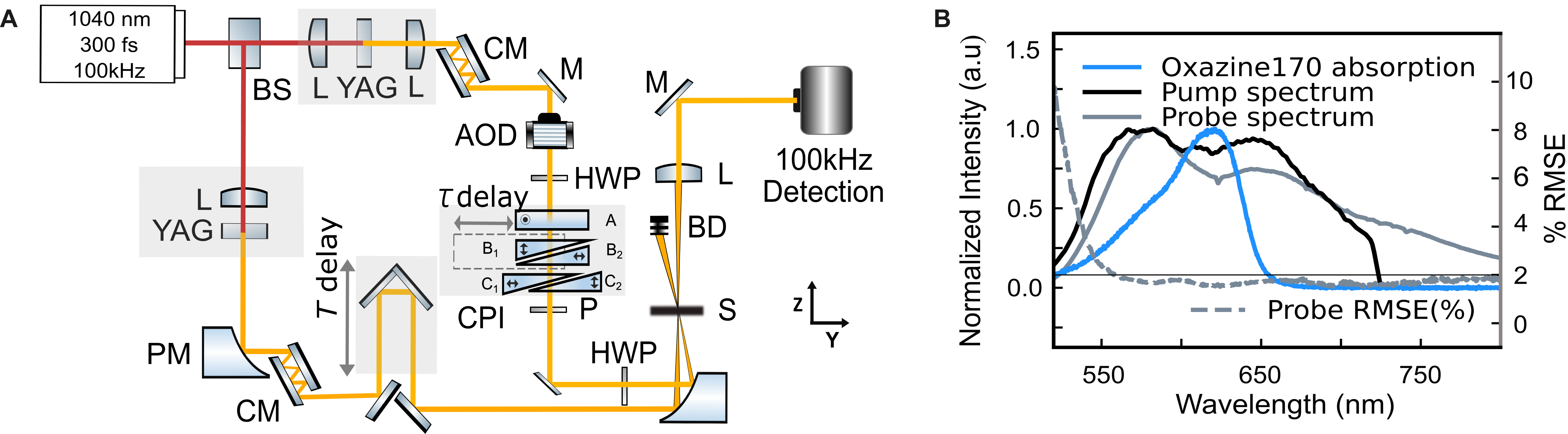}
	\caption{(a) Schematic of the WLC-2DES setup. BS Beam Splitter; L Lens; CM Chirped Mirror; M Mirror; AOD Acousto-Optic Deflector; HWP Half-waveplate; CPI Common Path Interferometer; P Linear Polarizer; PM Parabolic Mirror; S Sample; BD Beam Dump. 100 kHz detection represents the spectrograph, the 100 kHz line camera and the timing electronics that enable shot-to-shot detection. Dimensions of the wedges in the CPI (L$\times$W$\times$H) in mm ; 25x(3.6-0.5)x20, with an apex angle ($\theta$) of 7.07$^\text{{o}}$. (b) Linear absorption spectrum of Oxazine170 in Methanol along with averaged pump and probe
spectra. \% RMSE for probe passing through methanol in 500 $\mu$m cuvette is plotted along the secondary Y axis. The horizontal line is drawn at 2$\%$ RMSE. The average probe $\%$ RMSE measured through methanol in the range of 550-700 nm is 1.7$\%$. }
	\label{fig:fig1}
\end{figure}

The schematic of the partially collinear white-light 2DES setup is shown in {Fig.~\ref{fig:fig1}(a)}. The 1040 nm fundamental beam from a 100 kHz Yb:KGW amplifier (Spirit One, Spectra-Physics) is split into pump and probe lines of $\sim$ 1 $\mu$J power each. The fundamental beam is focused using 7.5 cm and 5 cm focal length lenses onto 8 mm and 10 mm YAG crystals for pump and probe WLC generation, respectively. Any residual of the fundamental is filtered using 725 nm (pump) and 850 nm (probe) shortpass optical filters (OD4, Edmund Optics). A crystalline quartz acousto-optic deflector (AOD, Gooch and Housego model 97-02965-01, 8.8 mm pathlength) modulates the pump at 50 kHz synchronously with the laser repetition rate ($f_R$) ensuring every other pump pulse is blocked. Note that the placement of AOD after the pump WLC generation introduces spatial chirp in the pump pulse. Placement of a collimation lens right after the AOD and reflective (achromatic) focusing \vt{with $\sim$33 $\mu$m average focal spot sizes} (substantially larger than $\sim$1 $\mu$m) are expected to mitigate\cite{Sahu2023} the effect of angular dispersion at the focus. However, measurements with sub-micron spatial resolution will necessarily require either double-passing\cite{Jefferts2005} through the AOD to exactly cancel out angular dispersion, or placement of AOD before the pump WLC generation. The deflected pump pulses are routed to a common path interferometer\cite{Cerullo2014} (CPI) for phase-locked pulse pair generation with mechanically controllable pump delay ($\tau$). More details of the CPI are described in Section \ref{cpi}. The total optical dispersion in the pump arm caused by the BBO wedges, optical filters, focusing and collimating lenses, sample cuvette, AOD and the YAG crystal is partially pre-compensated by two pairs of group delay dispersion (GDD) oscillation compensated chirped mirrors (Layertec 148545, -40 fs$^2$ GDD per bounce) with total 43 pairs of bounces where each bounce pair is specified to compensate $\sim$1 mm of fused silica. The probe beam is routed to the sample position after 22 pairs of bounces in a pair of chirped mirrors (148545 Layertec, -40 fs$^2$ GDD per bounce) to approximately compensate for optical dispersion in the probe WLC. \vt{A pump pulse duration of $\sim$33 fs is measured at the sample position by focusing into a SiC photodiode (Fig.~S1) and measuring the two-photon interferometric autocorrelation, and suggests uncompensated third-order or higher optical dispersion. The relaxed 2DES absorptive spectra reported here are not affected by these limitations of dispersion compensation with passive optical elements. The instrument response function (IRF) is assumed to be Gaussian and estimated to be \si{$\sim$60 fs} from a global fit of the rise time of the transient absorption signal measured with Oxazine 170} (\si{Fig.~S2}).\\

The delay ($T$) between the pump and probe arms is varied by a linear translational stage (ILS150BPP, Newport, 1 $\mu$m resolution). The pump and probe delay stages are controlled by a stage controller (XPS-D, Newport). The pump and probe arms with parallel polarization are focused into the sample in a 500 $\mu$m pathlength cuvette using a parabolic mirror (reflected focal length 101.6 mm) at a crossing angle of $\sim$7.5$^\text{o}$. The pump is blocked after the sample. The transmitted probe is routed to a spectrograph (Horiba iHR320, 150 grooves/mm) using a combination of reflective and achromatic optics. Every dispersed probe shot is recorded by a line CCD camera (e2v AViiVA, 14×28 $\mu$m, 1024 pixels) attached to the spectrograph. The CCD camera is interfaced with an Xtium-CL MX4 frame grabber (512 MB onboard memory buffer). The averaged pump and probe spectra are shown in {Fig.~\ref{fig:fig1}(b)} along with the \% root-mean-square (RMS) noise obtained by averaging 2k probe shots after transmission through the solvent. The pump and probe 1/$e^2$ focal spot sizes at the sample location were measured to be 33 $\mu$m and 36 $\mu$m, respectively (Fig.~S3(B-C)), with pulse energies \vt{0.47 nJ} and 1.53 nJ across the entire $>$150 nm WLC bandwidth. The sample \% transmission was confirmed to be linear across this range of pulse energies. Before each experiment, the overlap of pump and probe focal spots inside the sample cuvette along the optical axis was further optimized by maximizing the pump-probe signal as the cuvette position is changed (\si{Fig.~S3(A)}). This becomes crucial\cite{Cho2013} in case of a combination of large crossing angles, high sample ODs and long sample pathlengths where the signal may be dominantly generated towards the front of the sample. For the 2DES measurements reported here, Oxazine 170 (Sigma-Aldrich) solution is prepared in methanol with OD $\sim$0.37 in 500 $\mu$m cuvette with subsequent dilutions to prepare solutions of lesser OD. The OD of the samples was measured before and after the 2DES experiments and showed no changes beyond the measurement errors.\\

\subsection{Common Path Interferometer (CPI)}\label{cpi}

The CPI in the pump line in \si{Fig.~\ref{fig:fig1}(a)} is essentially a Babinet-Soleil compensator, and its design and application in 2DES presented here is motivated\cite{Cerullo2014} from the extensive work of Cerullo and co-workers. The interferometer consists of a rectangular block A of the negative birefringent material $\alpha$-BBO with the fast optical axis oriented along the X direction (according to the coordinate axes defined in \si{Fig.~\ref{fig:fig1}(a))}. 

When a 45$^o$ polarized pulse passes through this block, the X polarized component ($V$) travels with a faster velocity with respect to the Y polarized component ($H$) resulting in a delay between the two polarization components. This is followed by blocks B and C, each comprised of two pairs of $\alpha$-BBO wedges assembled in the form of rectangular blocks. The orientations of the optical axes in each wedge pair are such that the $H$ component travels faster in one set of wedges (in B$_2$ and C$_1$) and both components travel with the same velocity in the other set of wedges (B$_1$ and C$_2$). This implies that the relative group delay (GD) between the $V$ and $H$ components can be precisely controlled by adjusting the relative thickness of block A ($d_A$) and the pathlength traveled in the wedges B$_2$ and C$_1$, $d_{B2}$ and $d_{C1}$, as GD($\lambda_o$) = GVM($\lambda_o$)$(d_A - d_{B2} - d_{C1})$. Here GVM($\lambda_o$) $= (v_{g,o}^{-1}(\lambda_o) - v_{g,e}^{-1}(\lambda_o))$, is the group velocity mismatch at the central wavelength $\lambda_o$ where the mismatch ultimately depends on the ordinary versus extraordinary refractive index of $\alpha$-BBO, denoted as `$o$' and `$e$' subscripts, respectively. \\

To scan the delay between the $H$ and $V$ components, the B pair of wedges is mounted on a motorized translational stage (MFA-CC, Newport, 0.1 $\mu$m resolution) which enables control over the thickness $d_{B2}$ of wedge B$_2$ in the pump path. However, the overall thickness of the medium (ideally) remains fixed during the scan since the wedges are mounted in the form of a rectangular block. The wedges in block C are static and at minimum insertion to correct for the pulse front tilt of the pulses emerging from block B. The collinear $H$ and $V$ components pass through an output linear polarizer ({LPVISC050-MP2, Thorlabs}) at 45$^o$ polarization to result in collinear pulses with a common polarization axis, followed by rotation to vertical polarization by an achromatic half-waveplate ({AHWP05M-600, Thorlabs}). The spectral resolution is determined by the maximum possible delay range ($\tau_{max}$) scanned by the interferometer, which is $\sim\pm$320 fs at $\lambda_o =~$620 nm resulting in a spectral resolution of \vt{$\sim${52} cm$^{-1}$} along the absorption axis $\omega_{\tau}$. Note that the spectral resolution is limited by the system due to fast optical dephasing along the optical coherence time $\tau$ at room temperature. A common path design ensures that relative timing jitters $\delta\tau$ between the pulses during a $\tau$ scan are naturally suppressed with interferometric stability\cite{Cerullo2014} (\si{Fig.~S4}). 

\begin{figure}[h!]
	\centering\includegraphics[width=10.15cm]{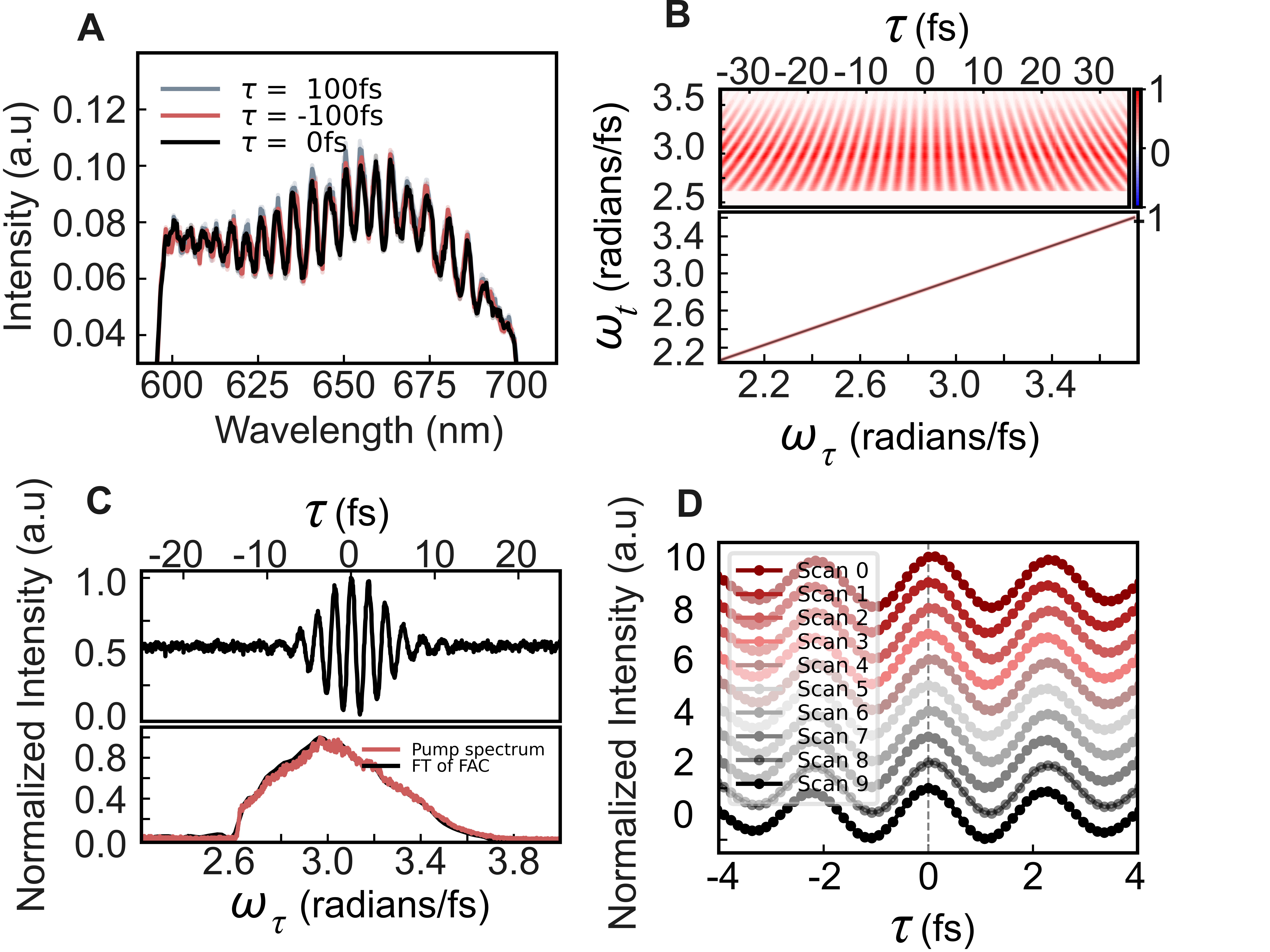}
	\caption{(a) Spectral \AST{interference} of the pump pulse $V$ and probe recorded at fixed $T$ delay for $\tau$ = -100, 0, 100 fs. Each spectrum is an average of five consecutive spectra at a fixed $\tau$. \vt{The error bar of the measurement across the five trials is overlaid on each spectrum.} (b) Top panel shows the spectrally-resolved Autocorrelation (sAC) of the pump pulses. Bottom panel shows the calibration of excitation frequency axis $\omega_{\tau}$ from sAC of the pump pulses by comparing the $\omega_{\tau}$ axis to the detection frequency $\omega_t$. The error bar in the calibration across measurements on consecutive days is overlaid as a red transparent band on the mean calibration curve. (c) Top panel shows the spectrally integrated sAC of the pump pulses recorded at the sample position zoomed into a range of $\pm$ 25 fs. Bottom panel compares the measured pump spectrum with the reconstructed pump spectrum obtained after Fourier transforming the trace. (d) Spectrally integrated sAC of pump pulses recorded by the shot-to-shot rapid scan detection scheme (Section \ref{sync}) for forward and backward scans. The scans are offsetted for clarity. Odd and even scans correspond to forward and backward scans, respectively \si{with a constant index shift between them (Fig.~S4(C))}. }
	\label{fig:fig2}
\end{figure}

Since the $V$ component travels through a fixed thickness of medium with constant group velocity irrespective of the stage position, the absolute time of emergence of the $V$ component should be ideally fixed during the $\tau$ delay scan of block B. This has been experimentally confirmed by focusing the $V$ component and the probe through a 25 $\mu$m pinhole and recording the spectral interference using a CCD spectrometer (CCS200, Thorlabs) as shown in \si{Fig.~\ref{fig:fig2}(a)}.  As the $\tau$ delay stage is scanned, the spectral fringes and the fringe density do not change confirming that $V$ component is unaffected during $\tau$ scan. This in turn ensures that the pump-probe waiting time $T$ remains constant during a $\tau$ scan. Note that during a $\tau$ scan, the $H$ component experiences a relative change in the thickness of ordinary and extraordinary glass whereas the $V$ component does not. This minor change in the amount of GDD between the two components is ignored in our analysis but can be consequential\cite{Cerullo2014} (in terms of relative pulse durations of the two pulses) for a combination of UV wavelengths, large $\tau$ scan range, and few cycle pulses. \si{Fig.~\ref{fig:fig2}(b)} (top panel) shows a spectrally-resolved autocorrelation sAC($\omega_t,\tau$) where the delay axis $\tau$ is constructed using the approximate conversion between stage position (scanned synchronously with the laser repetition rate $f_R$), insertion of block B and the resulting GD at $\lambda_o = $620 nm. Even though the stage position is synchronized to $f_R$, such an estimate is only approximate because the pulse replicas do not travel identical pathlengths in block B due to a small but finite air gap between the wedges B1 and B2 and a change in the fast axis orientation between the wedges. An exact calibration for the Fourier transformed frequency axis $\omega_{\tau}$ is obtained by comparing it to the detection frequency axis $\omega_t$ as shown in \si{Fig.~\ref{fig:fig2}(b)} (bottom panel). Prior to every experiment, sAC($\omega_t,\tau$) between the pump pulses is recorded with the shot-to-shot rapid scan detection scheme described later in Section \ref{sync}. The resulting calibration is checked prior to every experiment and was observed to be fairly consistent between day-to-day measurements as shown by the error bar in \si{Fig.~\ref{fig:fig2}(b)} (bottom panel). \\

\si{Fig.~\ref{fig:fig2}(c)} (top panel) shows the spectrally integrated sAC recorded at the sample position. The corresponding modulation depth, given by $\frac{I_{max}- I_{min}}{I_{max}+ I_{min}}\times$100, is calculated to be $\sim$ 93\%. A perfect modulation depth is expected for an ideal interferometer with perfect spatial overlap and collinearity between pulses. A deviation from 100$\%$ is likely caused by a finite air gap and a change in refractive index between the wedges due to which a few microns of lateral shift (relative to a 2 mm spot diameter) between the ideally overlapped and collinear $H,V$ components is expected\cite{Cerullo2014}. \si{Fig.~\ref{fig:fig2}(c)} (bottom panel) shows the comparison between the Fourier transform of the spectrally integrated signal in the top panel (reconstructed pump spectrum) and the measured pump spectrum. The calibration in \si{Fig.~\ref{fig:fig2}(b)} (bottom panel) was used to obtain the frequency axis of the reconstructed pump spectrum. The good agreement between the spectra confirms the validity of the excitation frequency axis calibration and the interferometer alignment and stability.\\

We have implemented the rapid scanning of the pump delay axis in both, forward and backward directions for faster averaging. In this regard, repeatability of $\tau$ points during consecutive scans is vital for efficient averaging of multiple $\tau$ scans without compromising the time step resolution. This is achieved by synchronizing the stage movement and CCD detection with the laser repetition rate $f_R$ as described in Section \ref{sync}. \si{Fig.~\ref{fig:fig2}(d)} shows the spectrally integrated sAC traces of the pump pulses recorded by the CCD line camera for consecutive forward and backward scans for the maximum stage velocity of \si{2 mm/s} used for the experiments reported here. Unlike previous implementations\cite{Cerullo2014}, synchronization of stage movement and CCD detection during $\tau$ scans ensures that all forward scans and all backward scans mutually overlap perfectly, with a constant index shift \si{(Fig.~S4(C))} between forward and backward scans which can be corrected during processing without recording separate pump interferograms. Without this synchronization, when multiple $\tau$ scans are averaged, arbitrary variations in $\tau$ points of the order of time steps are expected\cite{Cerullo2014}. Such variations can lead to phasing errors which become severe for a combination of faster scan velocities and lower repetition rates. The checks described in \si{Fig.~\ref{fig:fig2}} are conducted prior to every experimental run to confirm interferometer alignment and calibration. 

\subsection{Electronic Synchronization Scheme}\label{sync}

\begin{figure}[h!]
\centering\includegraphics[width=15cm]{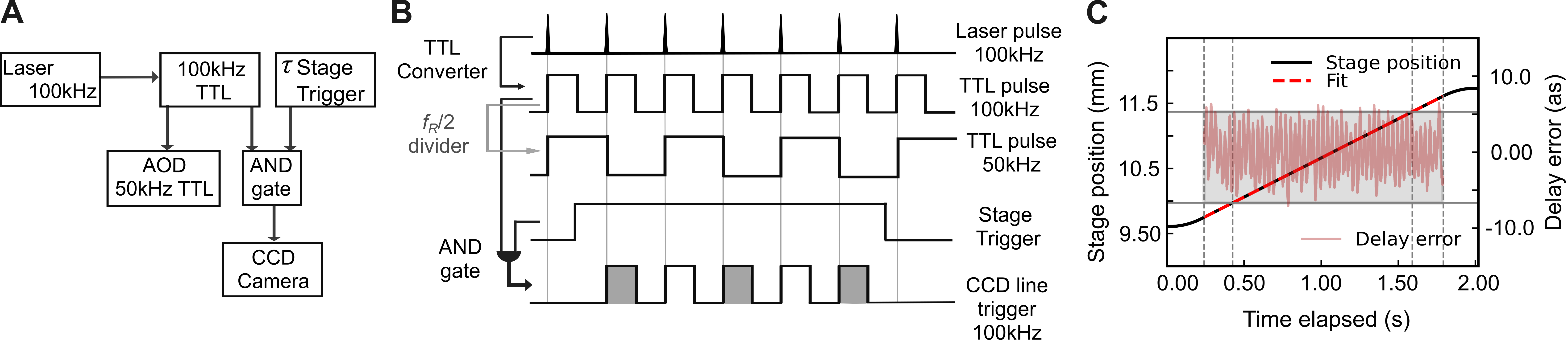}
\caption{(a) Schematic of the electronics synchronization for shot-to-shot data acquisition along with (b) the timing diagram. Such a detection scheme synchronizes the AOD, pump delay stage and the CCD camera to the laser repetition rate $f_R$. (Un)shaded region in the 100 kHz CCD line trigger corresponds to pump (ON)OFF state, while the probe is ON. (c) Motion profile of the $\tau$ delay stage recorded by the stage encoder for the \si{velocity of 1.2 mm/s}. The stage position versus time elapsed is plotted with the corresponding linear fit with a slope of \si{1.20 ± 3E{-6} mm/s compared to the set velocity of 1.2 mm/s.} The light red trace shows the corresponding delay error in attoseconds along the secondary Y axis. The vertical lines on the extreme ends correspond to the time window where the stage maintains 99.8$\%$ of the set velocity. The inner vertical lines define the trigger window over which the data is collected. The mechanical shutter is open from the start of the stage motion to the end of the trigger window. This window is defined as the sample exposure window. Note that in the motion profile, the stage is set to move 3$\times$ the calculated distance required by the stage to reach a constant velocity \vt{even though the encoder data suggests that the stage is already at 99.8\% of set velocity earlier than that}.}
\label{fig:fig3}
\end{figure}

\si{Figure \ref{fig:fig3}} (a-b) describes the timing electronics which synchronizes the laser repetition rate ($f_R$), pump modulation by AOD, $\tau$ stage movement and the CCD detection to combine shot-to-shot data acquisition with rapid delay scanning for any given input repetition rate. We have recently\cite{Bhat2023} implemented this detection scheme to demonstrate WLC-PP spectroscopy combining shot-to-shot detection at 100 kHz with continuous scanning of pump-probe waiting time $T$. Here we have extended this approach to WLC-2DES spectroscopy. The 100 kHz pulse train from the laser is converted to a 100 kHz TTL signal. The 100 kHz TTL output is then split into two parts of which one part is converted to a 50 kHz TTL signal by an $f_R$/2 TTL divider. This signal is used to drive the AOD such that every other pump shot is deflected into the setup, that is, pump modulation at $f_R$/2.  The pump delay stage controller outputs a constant high signal when the stage enters a defined $\tau$ scan window \AST{(trigger window)}. This signal is combined with the 100 kHz TTL pulse using an AND circuit, and the output is used to trigger the CCD camera. This results in the CCD camera reading every probe shot at repetition rate $f_R$ once the stage has entered the defined scan range. Furthermore, for every probe shot, the pump is alternating between $ON$ and $OFF$ states at $f_R$/2. In the pump-probe geometry, when the pump is $ON$, the 3$^{rd}$ order nonlinear signal is radiated in the same direction as the probe. The desired homodyned signal in case of shot-to-shot detection \vt{can be written}\cite{Jonas1999} as $S_{2D}$($\tau$, $T$, $\lambda_t$) = $S_{i+1}$($\tau$, $T$, $\lambda_t$)$^{ON}$ - $S_{i}$($\tau$, $T$, $\lambda_t$)$^{OFF}$ , where $S_{i+1}$ and $S_{i}$ denote consecutive transmitted probe shots recorded by the CCD with pump $ON$ and $OFF$, respectively. The homodyned signal is optically Fourier transformed by the spectrograph resulting in the detection wavelength axis ($\lambda_t$), which is then converted to a detection frequency axis ($\omega_t$) after a wavelength to frequency conversion during data processing. A numerical Fourier transform along the pump optical delay $\tau$ yields the absorption frequency axis $\omega_{\tau}$, to result in the absorptive 2D spectrum $S_{2D}$($\omega_\tau$, $T$, $\omega_t$) for a given pump-probe waiting time $T$. The signal has a transient absorption background $S_{TA}$($T$, $\omega_t$), which is constant along the $\tau$ delay axis for a fixed $\omega_t$ for a 2D spectrum measured at a fixed $T$. This constant offset can either be subtracted in the $\tau$ domain during data processing or removed by Fourier filtering in the $\omega_{\tau}$ domain. Note that in our implementation, we did not encounter the complication of alternating dark count background that is reported\cite{Grumstrup2019} to lead to dark count differences of the order of 1-5 counts between alternating lines in a PP microscopy experiment. Fourier filtering along $\tau$ implies that such a complication will not affect 2DES.

The rapid scan of the pump delay axis in principle leads to shot-to-shot increment of the pump delay with the theoretically expected minimum delay step given by $\Delta\tau_{min} = \text{GVM}(\frac{2v_{scan}}{f_R})\tan\theta$, where $\theta$ is the apex angle of the $\alpha$-BBO wedges (\si{Fig.~\ref{fig:fig1}}). This corresponds to the stage movement during two consecutive pump shots and is determined by the laser repetition rate $f_R$ and the stage velocity $v_{scan}$. Note that a slight timing offset between the stage trigger onset and the laser TTL high state (\si{Fig.~\ref{fig:fig3}(b)}) can lead to timing errors. However, the maximum possible such error ($\delta\tau$) encountered in a delay scan is given by $\delta\tau$  = $\text{GVM}(\frac{v_{scan}}{f_R})\tan\theta$. This error corresponds to  $\sim$1.1 attoseconds (as) for the maximum velocity of 2 mm/s implemented here. In our implementation this error is inconsequential because, as described in the following Section \ref{acq}, multiple finely sampled time steps, $\Delta\tau_{min}$, are binned together to form a larger $\tau$ steps of $\Delta\tau_{bin}$ such that the timing error is only $\sim$0.9\% of the binned delay step of 0.132 fs. Further details of shot-to-shot data acquisition and processing are described in Section \ref{acq}.

\begin{table}
\newcolumntype{p}[1]{>{\centering\let\newline\\\arraybackslash\hspace{3.5pt}}m{#1}}
\caption{Scan parameters for different stage velocities. The $\tau$ scan range of \vt{75.910 fs (-5.608 to 70.302 fs)}, binned time step $\Delta\tau_{bin}$ = 0.132 fs with \vt{15 points per 620 nm cycle} and $S$ = 10 scans are kept fixed across all experiments. The trigger window is defined in Fig.~\ref{fig:fig3}(c). The sample exposure window is also defined in Fig.~\ref{fig:fig3}(c) and includes 30 ms each for shutter opening and closing. A frame is defined as 1000 probe shots. The dead time is defined as the writing time after each $\tau$ scan and may not be needed for faster scan velocities where the number of frames collected per scan is substantially lesser than the frame grabber onboard memory.}

\centering
\footnotesize
\begin{tabular}{p{0.71in} p{0.71in} p{0.71in} p{0.74in} p{0.72in} p{0.69in} p{0.76in} p{0.82in}}
	\hline
	Stage velocity (mm/s) & \vt{Exposure Window} (secs) & Trigger Window (secs) & $ON/OFF$ pairs per bin ($M$) & Theoretical $\Delta\tau_{min}$ (as) & $\Delta\tau_{bin}$ (fs) & Frames Recorded per $\tau$ Scan & Dead Time per $\tau$ scan (secs)  \\ \hline
	$v$$_1$ = 0.4   & 4.192   & 3.498  & 300  &  0.5  &  0.132  & 343  & 7.642 \\
	$v$$_2$ = 0.8   & 2.337   & 1.748  & 150  &  0.9  &  0.132  & 171  & 3.840\\
	$v$$_3$ = 1.2   & 1.693   & 1.165  & 100  &  1.4  &  0.132  & 114  & 2.481 \\
	$v$$_4$ = 2.0   & 1.221   & 0.698  & 60    &  2.2  &  0.132  & 68    & 1.628 \\ \hline
	
\end{tabular}\label{Table1}
\end{table}
\normalsize

%

\subsection{Data Acquisition and Averaging Scheme}\label{acq}

Similar to rapid scan of $T$ delay stage at a fixed velocity within the defined stage trigger window in a PP experiment\cite{Bhat2023}, here the $\tau$ stage moves at a constant velocity within the defined trigger window. This is ensured by allowing the stage to move $\sim$3$\times$ the calculated distance $d$ required by the stage to accelerate to, or decelerate from, a uniform velocity before and after the defined trigger window, respectively. The set final velocity, set acceleration, the resulting distance $d$ and the time required by the stage to attain constant velocity is summarized in \si{Table S1}. The resulting motion profile of the stage as recorded by the stage encoder is shown in \vt{Fig.~\ref{fig:fig3}(c) for a representative stage velocity of 1.2 mm/s}. The region enclosed by the outer vertical lines corresponds to the region over which the stage moves with constant velocity. The shaded region enclosed by the inner set of dashed vertical lines represents the defined trigger window in which the probe shots are recorded whereas the outer most dashed vertical lines represent the region in which the stage velocity reaches 99.8\% of the set velocity. The sample is exposed to light from start of the $\tau$ stage motion to the end of the trigger window, after which a mechanical shutter blocks the light. As shown in the figure, delay errors are estimated by measuring the stage position deviations compared to that expected from a perfectly uniform motion profile, that is, a line with a constant slope corresponding to the set velocity. The resulting $\tau$ delay errors are $\sim$5 as, which is only 3.8\% of the binned time step $\Delta\tau_{bin}$ and of the order of $\sim$0.1 $\mu$m of wedge insertion (corresponding to the minimum incremental resolution of the stage).

\begin{figure}[hbt!]
\centering\includegraphics[width=11cm]{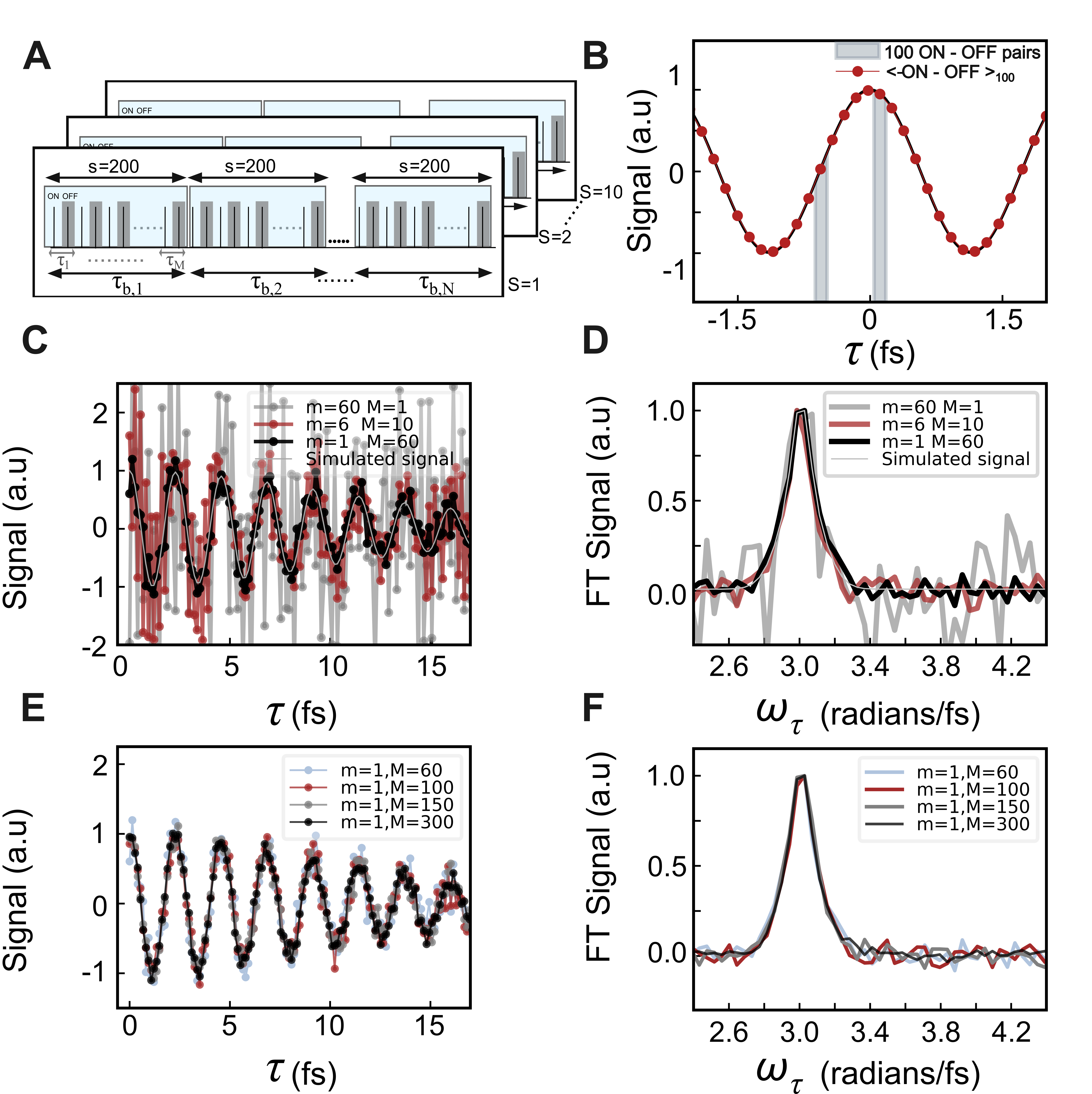}
\caption{Comparison of different averaging schemes. (a) Data averaging scheme shown for the case of $v$$_3$ scan in Table \ref{Table1}.  Black vertical lines represent consecutive probe shots. (Un)shaded represents probe shots recorded with pump $(ON)OFF$ state. Blue box represents the number of consecutive $ON/OFF$ pairs ($s$ probe shots and $M$ = $s$/2 pairs) averaged together to form one binned delay point $\tau_{b,i}$. $S$ represents one complete $\tau$ scan. Several such scans $S$ are conducted for each pump-probe delay point $T$. (b) Schematic representation of binning for the data collection in panel A. Simulated signal sampled with $\Delta\tau_{min}$ (black) and with $M$ = 100 averaged pairs constituting one binned point $\tau_{b,i}$, denoted as red dot. The area under the curve for 100 $ON/OFF$ pairs is denoted by shaded gray region around the binned data point and corresponds to 0.132 fs on the binned delay axis. (c) Experimentally measured shot-to-shot probe intensity fluctuations added to the simulated signal $S$$_{2D}$($\tau_{b,i}$) at $T$ = 1 ps and $\lambda_t$ = 645 nm where the signal maximizes. Three different averaging schemes are compared. $ON/OFF$ pairs with $m$ shots $ON$, $m$ shots $OFF$ and with total $M$ such $ON/OFF$ pairs are averaged together to yield one binned point $\tau_{b,i}$. $m$ = 1 is the shot-to-shot detection case with $M$ = 60 pairs averaged in case of scan velocity $v_4$ (Table \ref{Table1}). The plot is a zoomed in version of the full 70 fs scan range. Note that ($m,M$) are chosen such that the total laser shots contributing to a binned time step $\Delta\tau_{bin}$ is fixed for all cases for a fair comparison. (d) Normalized Fourier transform of the data shown in panel C. (e) Simulated signal with experimental probe noise for the shot-to-shot ($m$ = 1) case, for four different number of $ON/OFF$ pairs ($M$) per $\tau_{b,i}$. This is done in order to simulate the effect of $M$, resulting from scanning with four different velocities (Table \ref{Table1}), on the experimental signal to noise.  (f) Normalized Fourier transform of the data shown panel E.}
\label{fig:fig4}
\end{figure}


\si{Figure \ref{fig:fig4}} explains the data averaging scheme. As shown in \si{Fig.~\ref{fig:fig4}(a)}, the synchronization of the AOD, $\tau$ stage and the detector read-out to the laser repetition rate enables recording every probe shot with alternate pump $ON/OFF$ state. The continuous motion of the $\tau$ stage during this detection enables synchronous increment of the pump delay between consecutive $ON/OFF$ pairs resulting in finely sampled signal along the $\tau$ axis with stepsize $\Delta\tau_{min}$. For data processing, $M$ such pairs are averaged together to yield the $i^{th}$ binned time point $\tau_{b,i}$. Thus, as defined in Fig.~\ref{fig:fig4}(a), each binned time point $\tau_{b,i}$ arises from $M = s/2$ $ON/OFF$ pairs, that is, $M$ finely sampled $\tau$ points per bin. For example, for a velocity of 1.2 mm/s, $M $= 100 points per bin. Fig.~\ref{fig:fig4}(b) schematically denotes this binning procedure for a zoomed-in simulated signal $S_{2D}$($\tau_{b,i}, T, \lambda_t)$ at a fixed $T$ and $\lambda_t$, where a red dot denotes a binned point $\tau_{b,i}$ and the width of gray bar denotes the number $M$ of finely sampled $\tau$ points that together constitute an averaged $\tau_{b,i}$ time point. In our experiments, the number of points per bin, $M$ is adjusted for each velocity so as to keep the binned time step of $\Delta\tau_{bin}\sim$0.132 fs, resulting in directly comparable frequency axis $\omega_{\tau}$ across all velocities. Each $\tau$ scan is repeated `$S$' times at a fixed pump-probe delay $T$, with the $\tau$ stage moving alternately in forward and backward directions for consecutive scans. Apart from averaging $M$ points per binned step, averaging multiple $\tau$ scans further suppresses the effect of 1/$f$ long term laser drifts. Our recent shot-to-shot rapid scan PP\cite{Bhat2023} measurements also suggest that increasing the number of scan averages $S$ is more effective in suppressing the low-frequency 1/$f$ experimental noise than an equivalent increase in the number of points $M$ per binned point $\tau_{b,i}$. The scan parameters for the experiments with different stage velocities in Section \ref{results} are summarized in \si{Table 1}. \vt{These include the sample exposure time, finely sampled $\tau$ step ($\Delta\tau_{min}$), number of $ON/OFF$ pairs per bin ($M$), etc}. The $\tau$ scan range in the experiments, binned time step $\Delta\tau_{bin}$ = 0.132 fs with \vt{15 points per 620 nm cycle} and $S$ = 10 scans are kept fixed across all experiments. \\

As shown in Table 1, for a fixed scan range and the repetition rate $f_R$, the number of probe shots recorded with pump $ON/OFF$ states depends on the stage velocity. Defining a frame as $s$ = 1000 consecutive probe shots, in the sequence acquisition\cite{imaq} mode implemented here the onboard memory of the frame grabber can store a maximum of 490 frames. We therefore write a 2D spectrum on the computer RAM after each $\tau$ scan, leading to a dead time after each scan during which the frames are dumped on the computer memory. However, for the faster velocities implemented here, the number of frames per $\tau$ scan are substantially lesser than maximum possible, for instance 68 frames per $\tau$ scan for 2 mm/s velocity. In such cases, a dead time after each $\tau$ scan is not required because all scans can be recorded together before writing them on the computer RAM. \\

Note that in Table \ref{Table1}, $\Delta\tau_{min}$ for even the fastest velocity corresponds to a wedge insertion $\sim$2.5$\times$ lesser than the minimum incremental resolution of the stage, implying that the maximum velocity that could be used in our approach can be at least 2.5$\times$ faster. However, we cannot use arbitrarily fast stage velocity because stage movement between pump $ON$ and $OFF$ states can no longer be ignored if the signals $S_{i+1}^{ON}$ and $S_{i}^{OFF}$ vary substantially with $\tau$ stage movement between consecutive shots. In this regard, programmable pulse shaping approaches hold a distinct advantage because of no mechanical delay elements. Consequently, in the AOPS approach\cite{Kearns2017} of Kearns et al. a `burst scan' approach is employed where instead of $M$ points per bin and $S$ scans, an equivalent of $M \times S$ scans with no binning can be implemented with a better suppression of 1/$f$ noise encountered over the duration\cite{Moon1993} of the $\tau$ scan. In terms of scan efficiency\cite{Bhat2023}, or the number of pulses utilized versus wasted for a measurement, AOPS approach holds a clear advantage as well, because no mechanical delay elements are involved and therefore no pulses are wasted while the stage attains a constant velocity or the scan direction is alternated between forward and backward.  Note that if scan efficiency is the main concern, Fig.~\ref{fig:fig3}(c) suggests that the requirement of 3$d$ distance can be easily relaxed as well as the time required to attain a constant velocity can be minimized by increasing the stage acceleration (Table S1). However, maximizing scan efficiency has no bearing on the 1/$f$ experimental noise encountered \textit{during} a scan. Despite these distinct advantages of programmable pulse shaping, the results in Section \ref{results} demonstrate that comparable throughput and SNR is attainable even with mechanical delay scans through a combination of shot-to-shot detection with the rapid scan approach, with vital advantages of simplicity and repetition-rate scalability without sacrificing pump WLC bandwidth.


\section{Results and Discussion}\label{results}

This section compares various averaging schemes to \vt{demonstrate advantages of shot-to-shot detection}, followed by rapid scan shot-to-shot 2DES experiments on Oxazine 170 including a demonstration of SNR for different stage velocities and sample concentrations.

\subsection{Comparison of Averaging Schemes}
\si{Fig.~\ref{fig:fig4}(c)} compares the SNR possible with various averaging schemes in order to motivate the 2DES experiments with rapid scan shot-to-shot detection. `$m$' denotes the number of probe shots that are averaged together with pump $ON$ or $OFF$. $m$ = 1 case results in the shot-to-shot 2D signal $S_{2D}$ defined earlier. In contrast, $m$ = 60 implies that 60 probe shots are averaged with pump $ON$ and $OFF$, and a difference of the two then yields $S_{2D}$.`$M$' is the number of such $ON/OFF$ pairs that are averaged together to result in a 2D signal at time $\tau_{b,i}$. The panel considers the case of maximum scan velocity $v_4$ in Table \ref{Table1}. Note that the total number of probe shots `$s$' that result in the averaged signal, $s$ = 2$m$$M$, is kept fixed for a fair comparison between the three averaging schemes. This implies that the reduction in 1/$f$ component of experimental noise over the duration of the $\tau$ scan \cite{Moon1993,Kearns2017} is equivalent between the three averaging schemes. The probe laser noise is estimated directly from experiment by passing the probe through the sample with pump blocked, recording every probe shot, and subtracting the mean counts. The probe noise at $\lambda_t = $ 645 nm, where the signal maximizes, is then grouped together as per the ($m$,$M$) combination and added to the simulated signal. \si{Fig.~\ref{fig:fig4}(d)} shows the corresponding Fourier transform. The analysis in Figs.\ref{fig:fig4}(c-d) demonstrates that faster the modulation of PP signal, better is the resulting SNR because pump-induced intensity modulations in the probe at the maximum possible frequency of 2/$f_R$ minimizes the 1/$f$ component\cite{Moon1993} of probe noise for a given $\tau$ data point. The relative standard deviation is quantified in Fig.~S5(A) and corresponds to  1x, 3.63x and 5.68x for ($m,M$) = (1,60), (6,10), and (60,1), respectively. 

Fig.~\ref{fig:fig4}(e-f) simulates the effect of the four different $\tau$ scan velocities in Table \ref{Table1} on the SNR, for the case of $m$ = 1, that is, shot-to-shot detection. As before, experimentally measured shot-to-shot probe noise is binned depending on the ($m,M$) = (1,$M$) combination, and added to the simulated signal. Number of points per bin, $M$ is chosen to be 300, 150, 100, 60 corresponding to the four velocities $v_1$--$v_4$ (Table \ref{Table1}). \si{Fig.~S5 (B) compares the standard deviation of the time-domain noise floor for $M$ = 60, 100 and 150 relative to the $M$ = 300 case, against that expected from $1/\sqrt{M}$ scaling of Gaussian noise with bin size. The noise floor in case of faster scans degrades lesser than that expected from simply scaling the bin size, suggesting advantages of shot-to-shot detection for faster scan velocities. This trend is qualitatively similar to the reported\cite{Kearns2017}  suppression of 1/$f$ noise encountered over the scan duration as the scans become faster. Overall, the SNR for the v$_4$ scan ($M$ = 60) is only $\sim$1.78$\times$ lesser but with an advantage of 5$\times$ faster acquisition suggesting equivalent advantages in the 2DES experiments.} 

Note that in addition to faster throughput, rapid delay scan also proportionally minimizes the sample exposure time which is crucial to optimize in order to mitigate sample photodamage in micro-spectroscopy applications. Sample exposure time per $\tau$ scan is tabulated in Table 1 with one 2DES spectrum collected in $\sim$1.2 seconds for the maximum velocity implemented here. Note that this exposure time includes the additional time required by the stage to reach from the point of 99.8\% of set velocity to the start of the trigger window (time between the initial two vertical lines in Fig.~\ref{fig:fig3}C), as well as the shutter opening and closing times before and after each $\tau$ scan. As discussed in Section \ref{acq}, future experiments without this additional time, and with at least 2.5$\times$ faster velocity can significantly minimize this exposure time in a feasible manner without running into issues related to stage movement during pump $ON$ and $OFF$ states in a rapid scan approach. Based on these simulations, Sections \ref{rapid2D}--\ref{sensitivity} present shot-to-shot 2DES spectra for the rapid scan settings in Table \ref{Table1} to demonstrate the throughput and sensitivity of the approach proposed here.

\subsection{Shot-to-shot Rapid Scan 2D Spectra}\label{rapid2D}
\begin{figure}[h!]
\centering\includegraphics[width=15.2cm]{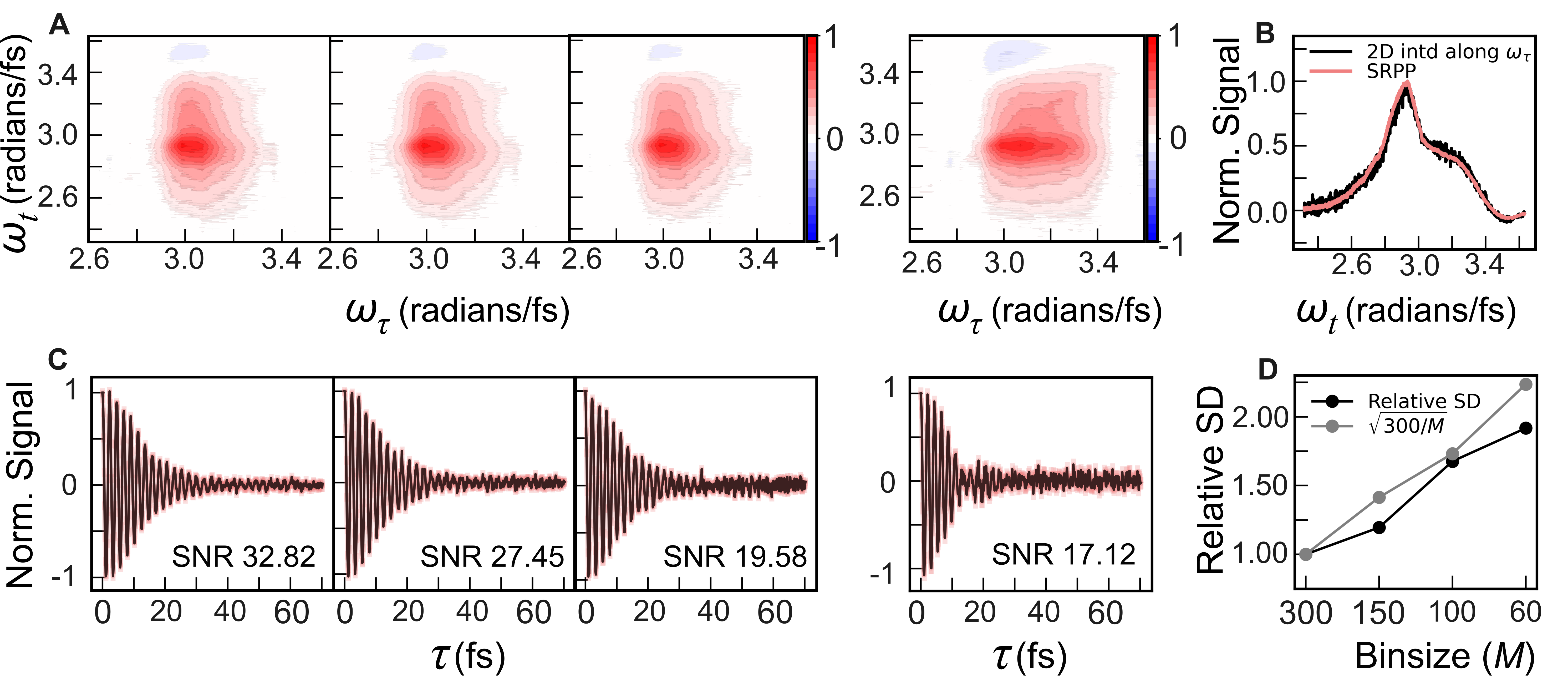}
\caption{(a) Normalized 2D spectra of 0.37 OD Oxazine 170 in Methanol at $T$ = 1 ps recorded with stage velocities $v_1-v_3$ and $v_4$. Contours are drawn at 5$\%$ and 10$\%$–100$\%$ in 10$\%$ intervals for both positive or negative contours. The 2D spectrum for the fastest velocity $v_4$ is broader and blue-shifted compared to other cases because this experiment was conducted on a different day with a blue shifted pump bandwidth (Fig.~S7). (b) Spectrally integrated 2D spectrum (along $\omega_{\tau}$) corresponding to $v_4$ overlaid on the spectrally-resolved pump-probe (SRPP) spectrum at $T$ = 1 ps. (c) $S$($\tau$) at $T$ = 1 ps, $\omega_t$ = 2.918 rad/fs for $v_1-v_3$ and $v_4$. The error bar calculated over $M$ bins per $\tau$ point and $S$ scans is overlaid as a translucent band on the signal. SNR calculated for each slice is mentioned in the inset. (d) Standard deviation versus the bin size for the different scan velocities. The standard deviation for each case is normalized relative to the $M$ = 300 ($v_1$) case. Gray trace overlays the curve expected from the $1/\sqrt M$ scaling with decreasing bin size $M$.}
\label{fig:fig5}
\end{figure}

\si{Figure~\ref{fig:fig5}} shows the absorptive 2D spectra $S_{2D}$($\tau$, $T$, $\omega_t$) at $T $= 1 ps collected for scan settings in Table \ref{Table1}. Experimental 2D data on Oxazine 170 in methanol are collected for a $\tau$ scan range of -5.608 to 70.302 fs for a fixed pump-probe delay ($T$). The constant offset between the forward and backward scans is aligned before the $\tau$ scans are averaged together. The averaged 2D data is spectrally integrated along the detection frequency axis and the maximum of the interferogram is determined. The spectrally integrated signal maximizes at zero $\tau$ delay between the pump pulses. The data is cropped at the maxima before the Fourier transform along the $\tau$ delay axis. Any error in determining the signal maximum results in phasing issues, that is, mixed absorptive and dispersive lineshapes along $\omega_{\tau}$ in the relaxed 2D spectrum. Minor phasing errors of the order of half the bin size can arise due to the binning procedure described in Section \ref{acq}. This is so because the binned data does not necessarily sample the signal at $\tau$ = 0. In other words, the maxima of the binned signal is at $\tau$ = 0 $\pm$ $\Delta$$\tau$$_{shift}$. This shift in delay results in an extra phase factor in the Fourier transformed signal and needs to be quantified to accurately phase the 2D spectra. This is done by comparing the binned $\tau$ scans with a finely binned scan with $M$ = 25, and correcting for $\Delta$$\tau$$_{shift}$ in the frequency domain. The maximum timing error in this method is given by 0.5$\times$$\Delta\tau_{bin}$ for $M$ = 25, which corresponds to phase error of less than $\lambda$/369 at 620 nm. This procedure is illustrated in Fig.~S6.

\si{Fig.~\ref{fig:fig5}(a)} shows the 2D spectra of Oxazine 170 at $T$ = 1 ps with sample OD 0.37 in a 500 $\mu$m pathlength cuvette for the four scan velocities. The scan velocities and related parameters are shown in Table \ref{Table1}. The frequency resolution along the $\tau$ axis is system limited due to fast optical dephasing. For a scan range of $\sim$70 fs, the resulting frequency resolution after maximum allowed $N$ to 2$N$ zero-padding is $\sim$238 cm$^{-1}$. The 2D spectrum for the fastest velocity is broader and blue-shifted compared to other cases because this experiment was conducted on a different day with a blue shifted pump bandwidth (Fig.~S7). The changes in the 2D spectra corresponding to different pump bandwidths compare well with those expected from the relaxed 2D spectrum constructed with the experimental pump and probe bandwidths, and independently measured absorption and spontaneous emission lineshapes\cite{Jonas2001} (Fig.~S7). \si{Fig.~\ref{fig:fig5}(b)} compares the spectrally integrated phased 2D spectrum at 1 ps with the spectrally-resolved pump probe spectrum at 1 ps (\si{Section S2}) for the fastest scan velocity. Overlap of the two spectra indicates no residual phase in the recorded signal along the detection axis as may be expected in homodyned detection \cite{Jonas1999}.

\si{Fig.~\ref{fig:fig5}(c)} shows the $\tau$-domain traces at  $\omega$$_t$ = 2.918 rad/fs (\si{$\lambda_t$ = 645 nm}) for the four stage velocities with the error bar overlaid. The SNR for each case is estimated similar to ref.\cite{Kearns2017}, by the inverse of the standard deviation of the normalized signal for a range of $\tau$ > 40 fs where the signal has completely dephased. \si{Fig.~\ref{fig:fig5}(d)} plots the corresponding standard deviation versus the bin size for the three velocities. The SNR is lowest for the fastest velocity $v_4$, as expected for the smallest bin size. However, $v_4$ reduces the number of probe shots needed to record a 2D spectrum by 5$\times$ with SNR deteriorating only by $\sim$1.9$\times$ compared to the slowest scan (maximum points per bin case). When this measured SNR in Fig.~\ref{fig:fig5}d is compared against the expected $1/\sqrt{M}$ dependence of SNR at each $\tau$ data point, the noise floor increases with decreasing bin size as expected. However, similar to the trend in the simulations in \si{Fig.~S5}, the noise floor consistently degrades lesser than that predicted by the $1/\sqrt{M}$ scaling with decreasing bin size $M$. This again emphasizes the point that rapid scan with shot-to-shot detection is expected to suppress\cite{Kearns2017} the low-frequency 1/$f$ noise encountered\cite{Moon1993} over the duration of a scan. Note that such a suppression will be maximum for a `burst' scan (Section \ref{acq}), which is straightforward in the AOPS approach because phase errors arising from stage movement during consecutive pump $ON$ and $OFF$ states (Section \ref{sync}) are entirely circumvented in programmable pulse shaping.\\

\subsection{Sensitivity}\label{sensitivity}

\begin{figure}[hbt!]
\centering\includegraphics[width=13.2cm]{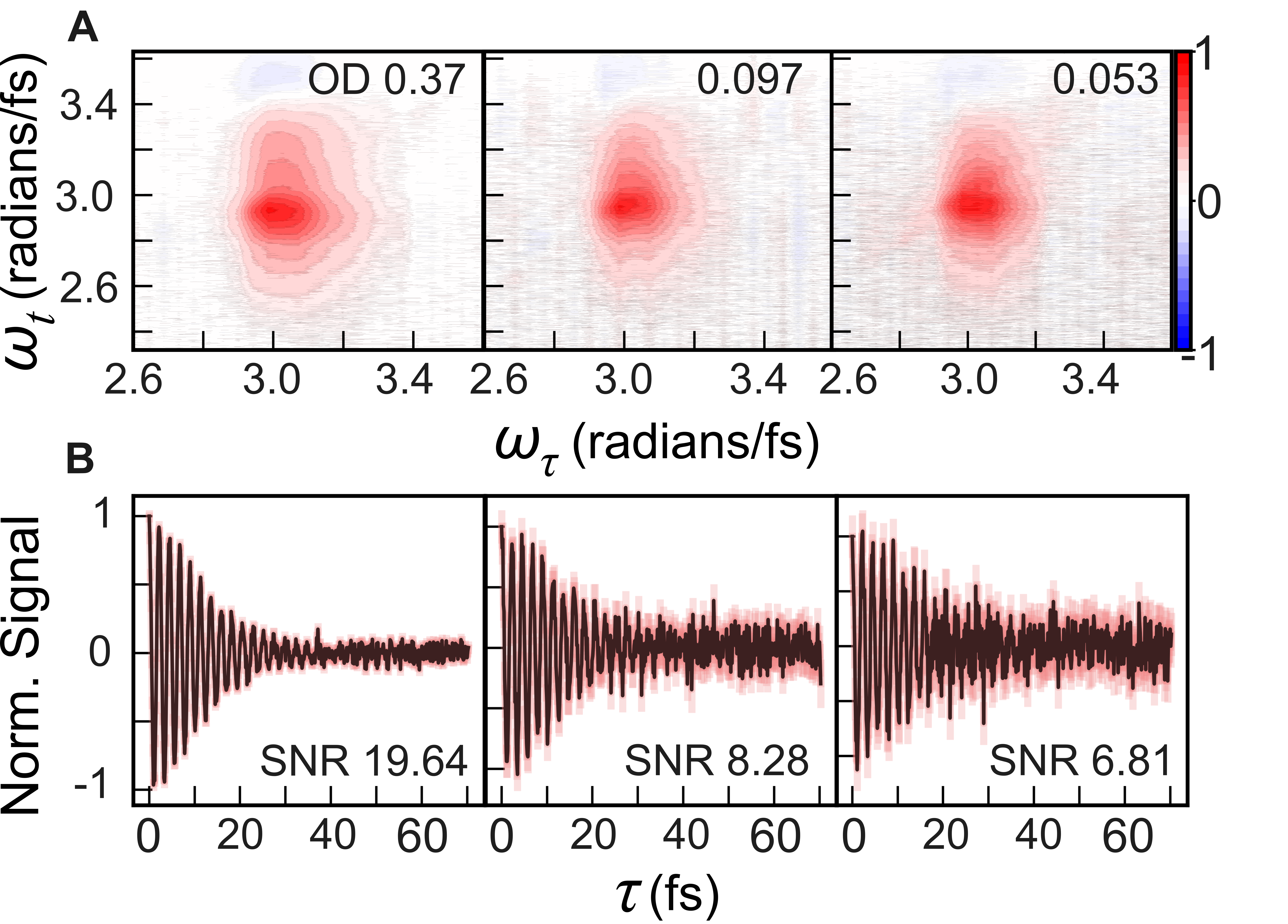}
\caption{(a).Normalized $T$ = 1 ps 2D spectra of Oxazine 170 at three different concentrations recorded at velocity $v$$_3$ . Contours are drawn at 2$\%$ 5$\%$ and
10$\%$–100$\%$ in 10$\%$ intervals for both positive or negative contours. (b) $S$( $\tau$ ,$T$, $\omega$$_t$) at $\omega$$_t$ = 2.918 rad/fs for the three concentrations. The error bar calculated over $M$ points per bin and $S$ scans averaged is overlaid as a translucent band on the signal. SNR calculated for each slice is mentioned in the inset.}
\label{fig:fig6}
\end{figure}

Encouraged by the SNR degradation lesser than what is expected from $1/\sqrt{M}$ scaling of Gaussian or random noise (Fig.~\ref{fig:fig5}(d)) along with simultaneous throughput improvement, we decided to test the sensitivity of rapid scan shot-to-shot detection approach by reducing the sample concentration at the velocity of 1.2 mm/s, that is, 3$\times$ faster than the case for which best the SNR is measured (due to more number of points per bin). The starting OD in 500 $\mu$m cuvette is 0.37 as before, which corresponds to a number density of 5.5E16 molecules/cm$^3$. Figure~\ref{fig:fig6}(a) shows consistent measurements of 2D spectra for concentrations down to 0.78E16 molecules/cm$^3$ corresponding to an OD of 0.053 in 500 $\mu$m cuvette. A small red-shift of $0.029$ rad/fs (6.3 nm) is seen along the detection axis from lowest to highest OD. We checked that this trend is consistent with the measurement of fluorescence spectra (Fig.~S8) which show a progressively increasing red-shift from lowest to highest OD likely caused by aggregation at higher concentrations. The SNR in Fig.~\ref{fig:fig6}(b) degrades with decreasing concentration as is expected. Degrading SNR implies that the negative 5\% excited state absorption (ESA) signal contour at $\omega_t$ = 3.52 rad/fs is at the noise floor. This level of SNR is achieved with total 1140E3 probe shots (Table \ref{Table1}), that is, $\sim$11.6 seconds of averaging. The total experimental time is $\sim$3$\times$ longer because it includes the full sample exposure window and the dead time which lowers the overall scan efficiency\cite{Bhat2023} in case of rapid scanning of mechanical delays as compared to the AOPS approach. As discussed in Section \ref{acq}, the rapid scan efficiency for a given velocity can be improved by avoiding dead times and minimizing the additional distance traveled by the stage at a constant velocity. Note however that the scan efficiency has no bearing on the SNR reported here which reports the SNR encountered \textit{during} a scan, and therefore does not depend on the dead times encountered before or after a scan. If desired, a higher level of sensitivity is also straightforward to achieve by using slower scan velocities, that is, larger bin size $M$ or longer averaging through slower scan velocity.\\

In comparison to above SNR and averaging times reported for sample concentrations of 0.78E16 molecules/cm$^3$, the state-of-the-art AOPS approach to WLC-2DES has reported\cite{Kearns2017} an SNR of 4.2 in 180 secs of averaging at 100 kHz for a sample concentration of 26E16 chlorophyll a molecules/cm$^3$ (OD 0.08 in 50 $\mu$m cuvette). Note that our crossing angle of 7.5$^o$ implies that pump-probe spot overlap will not perfect throughout the cuvette pathlength. This is further verified from Fig.~S3 where the maximum signal drops to approximately half within a 200 $\mu$m region as the cuvette is translated along the beam propagation direction. This suggests that the sample pathlength over which the pump-probe signal is predominantly generated is lesser than the cuvette pathlength and therefore the effective sample OD which generates the pump-probe signal may be lesser than 0.05. Note also that in comparison to the conventional 2DES approaches mentioned above, a recent rapid scan fluorescence-detection 2DES approach\cite{Sahu2023} has reported measurements of coherent 2D signals which are only $\sim$10\% of population signals for sample ODs as low as $\sim$1 mOD, although at 1 MHz repetition rate and with significantly longer averaging times.

\section{Conclusions}

We have introduced a repetition rate scalable approach to 2DES spectroscopy that combines the benefits of shot-to-shot detection with rapid scanning of mechanical delays to provide a viable alternative to state-of-the-art AOPS approaches.  Our approach relies on the simplicity of conventional optical elements to generate phase-locked pump pulse pairs and a broadband white light continuum as input. We demonstrate this through mutual synchronization between the laser repetition rate, acousto-optical deflector (AOD), pump delay stage and the CCD line camera, which allows rapid scanning of pump optical delay synchronously with the laser repetition rate while the delay stage is moved at a constant velocity. The resulting shot-to-shot detection scheme is repetition rate scalable with the throughput only limited by the CCD line rate and \vt{the maximum stage velocity} without any limitations imposed on the pump WLC bandwidth as $f_R$ increases beyond 100 kHz. Using this approach, we demonstrate measurement of an averaged 2DES absorptive spectrum in as much as \vt{1.2 seconds of continuous sample exposure per 2D spectrum}. We achieve a signal-to-noise ratio (SNR) of 6.8 for optical densities down to 0.05 with 11.6 seconds of averaging at 100 kHz laser repetition rate. We discuss limitations of mechanical delays compared to programmable pulse shaping in terms of `burst scans', where AOPS approaches can provide maximum 1/$f$ noise suppression and better scan efficiency. However, compared to AOPS approaches, the approach proposed here does not run into fundamental limitations of AOPS approaches at higher repetition rates such as limited time aperture-RF bandwidth product and RF update rate. Overall, combining rapid scan with shot-to-shot detection as demonstrated here provides throughput and sensitivity comparable to the AOPS approach, is repetition-rate scalable and minimizes sample exposure per 2D spectrum. Our demonstration opens door to promising micro-spectroscopy applications using a combination of repetition rate tunable Yb:KGW amplifiers and currently available cameras of up to 250 kHz line rates, which can be easily accommodated without any change in the experimental setup except the input TTL signal. 

%
%

%

\section*{SUPPLEMENTARY MATERIAL}

See the \si{supplementary material} for pulse width and instrument response function measurements, SRPP spectra, signal vs cuvette position and spot size measurements, phasing procedure, linear absorption and emission spectra of Oxazine170, and estimation of relaxed 2D spectrum from absorption and emission lineshapes. 

\section*{Funding}

This work is supported in parts by research grants from the Indian Space Research Organization (ISTC/CSS/VT/468); Department of Biotechnology, India (BT/PR38464/BRB/10/1893/2020); Board of Research in Nuclear Sciences (58/20/31/2019-BRNS) and the Science and Engineering Research Board (CRG/2019/003691, CRG/2022/004523).

\section*{Acknowledgments}

We thank Prof. Giulio Cerullo (Politecnico di Milano) for details of the specifications of birefringent wedges, and Prof. Minjung Son (Boston University) for initial suggestions regarding CCD line rate cameras. AST acknowledges Prime Minister's Research Fellowship, MoE India. VNB acknowledges research fellowship from DST-Inspire. VT acknowledges the Infosys Young Investigator Fellowship supported by the Infosys Foundation, Bangalore.

\section*{AUTHOR DECLARATIONS}
\subsection*{Conflict of Interest}

The authors have no conflicts to disclose.

\subsection*{Author Contributions}
VT designed the research problem. AST and VNB contributed equally to the work. All authors contributed towards writing the manuscript.

\section*{Data Availability}

Data underlying the results presented in this paper are not publicly available at this time but may be obtained from the authors upon reasonable request.

\bibliography{rapid2D}

\begin{thebibliography}{41}%
\makeatletter
\providecommand \@ifxundefined [1]{%
 \@ifx{#1\undefined}
}%
\providecommand \@ifnum [1]{%
 \ifnum #1\expandafter \@firstoftwo
 \else \expandafter \@secondoftwo
 \fi
}%
\providecommand \@ifx [1]{%
 \ifx #1\expandafter \@firstoftwo
 \else \expandafter \@secondoftwo
 \fi
}%
\providecommand \natexlab [1]{#1}%
\providecommand \enquote  [1]{``#1''}%
\providecommand \bibnamefont  [1]{#1}%
\providecommand \bibfnamefont [1]{#1}%
\providecommand \citenamefont [1]{#1}%
\providecommand \href@noop [0]{\@secondoftwo}%
\providecommand \href [0]{\begingroup \@sanitize@url \@href}%
\providecommand \@href[1]{\@@startlink{#1}\@@href}%
\providecommand \@@href[1]{\endgroup#1\@@endlink}%
\providecommand \@sanitize@url [0]{\catcode `\\12\catcode `\$12\catcode
  `\&12\catcode `\#12\catcode `\^12\catcode `\_12\catcode `\%12\relax}%
\providecommand \@@startlink[1]{}%
\providecommand \@@endlink[0]{}%
\providecommand \url  [0]{\begingroup\@sanitize@url \@url }%
\providecommand \@url [1]{\endgroup\@href {#1}{\urlprefix }}%
\providecommand \urlprefix  [0]{URL }%
\providecommand \Eprint [0]{\href }%
\providecommand \doibase [0]{https://doi.org/}%
\providecommand \selectlanguage [0]{\@gobble}%
\providecommand \bibinfo  [0]{\@secondoftwo}%
\providecommand \bibfield  [0]{\@secondoftwo}%
\providecommand \translation [1]{[#1]}%
\providecommand \BibitemOpen [0]{}%
\providecommand \bibitemStop [0]{}%
\providecommand \bibitemNoStop [0]{.\EOS\space}%
\providecommand \EOS [0]{\spacefactor3000\relax}%
\providecommand \BibitemShut  [1]{\csname bibitem#1\endcsname}%
\let\auto@bib@innerbib\@empty
\bibitem [{\citenamefont {Schoenlein}\ \emph {et~al.}(1991)\citenamefont
  {Schoenlein}, \citenamefont {Peteanu}, \citenamefont {Mathies},\ and\
  \citenamefont {Shank}}]{Mathies1991}%
  \BibitemOpen
  \bibfield  {author} {\bibinfo {author} {\bibfnamefont {R.~W.}\ \bibnamefont
  {Schoenlein}}, \bibinfo {author} {\bibfnamefont {L.~A.}\ \bibnamefont
  {Peteanu}}, \bibinfo {author} {\bibfnamefont {R.~A.}\ \bibnamefont
  {Mathies}},\ and\ \bibinfo {author} {\bibfnamefont {C.~V.}\ \bibnamefont
  {Shank}},\ }\bibfield  {title} {\enquote {\bibinfo {title} {{The first step
  in vision: femtosecond isomerization of rhodopsin}},}\ }\href
  {https://doi.org/10.1126/science.1925597} {\bibfield  {journal} {\bibinfo
  {journal} {Science}\ }\textbf {\bibinfo {volume} {254}},\ \bibinfo {pages}
  {412--415} (\bibinfo {year} {1991})}\BibitemShut {NoStop}%
\bibitem [{\citenamefont {Richter}\ \emph {et~al.}(2017)\citenamefont
  {Richter}, \citenamefont {Branchi}, \citenamefont {{de Almeida Camargo}},
  \citenamefont {Zhao}, \citenamefont {Friend}, \citenamefont {Cerullo},\ and\
  \citenamefont {Deschler}}]{Richter2017}%
  \BibitemOpen
  \bibfield  {author} {\bibinfo {author} {\bibfnamefont {J.~M.}\ \bibnamefont
  {Richter}}, \bibinfo {author} {\bibfnamefont {F.}~\bibnamefont {Branchi}},
  \bibinfo {author} {\bibfnamefont {F.}~\bibnamefont {{de Almeida Camargo}}},
  \bibinfo {author} {\bibfnamefont {B.}~\bibnamefont {Zhao}}, \bibinfo {author}
  {\bibfnamefont {R.~H.}\ \bibnamefont {Friend}}, \bibinfo {author}
  {\bibfnamefont {G.}~\bibnamefont {Cerullo}},\ and\ \bibinfo {author}
  {\bibfnamefont {F.}~\bibnamefont {Deschler}},\ }\bibfield  {title} {\enquote
  {\bibinfo {title} {{Ultrafast carrier thermalization in lead iodide
  perovskite probed with two-dimensional electronic spectroscopy}},}\ }\href
  {https://doi.org/10.1038/s41467-017-00546-z} {\bibfield  {journal} {\bibinfo
  {journal} {Nature Communications}\ }\textbf {\bibinfo {volume} {8}},\
  \bibinfo {pages} {376} (\bibinfo {year} {2017})}\BibitemShut {NoStop}%
\bibitem [{\citenamefont {Lang}(2018)}]{Lang2018}%
  \BibitemOpen
  \bibfield  {author} {\bibinfo {author} {\bibfnamefont {B.}~\bibnamefont
  {Lang}},\ }\bibfield  {title} {\enquote {\bibinfo {title} {{Photometrics of
  ultrafast and fast broadband electronic transient absorption spectroscopy:
  State of the art}},}\ }\href {https://doi.org/10.1063/1.5039457} {\bibfield
  {journal} {\bibinfo  {journal} {Review of Scientific Instruments}\ }\textbf
  {\bibinfo {volume} {89}},\ \bibinfo {pages} {93112} (\bibinfo {year}
  {2018})}\BibitemShut {NoStop}%
\bibitem [{\citenamefont {Dostál}, \citenamefont {Benešová},\ and\
  \citenamefont {Brixner}(2016)}]{Dostal2016}%
  \BibitemOpen
  \bibfield  {author} {\bibinfo {author} {\bibfnamefont {J.}~\bibnamefont
  {Dostál}}, \bibinfo {author} {\bibfnamefont {B.}~\bibnamefont
  {Benešová}},\ and\ \bibinfo {author} {\bibfnamefont {T.}~\bibnamefont
  {Brixner}},\ }\bibfield  {title} {\enquote {\bibinfo {title}
  {{Two-dimensional electronic spectroscopy can fully characterize the
  population transfer in molecular systems}},}\ }\href
  {https://doi.org/10.1063/1.4962577} {\bibfield  {journal} {\bibinfo
  {journal} {The Journal of Chemical Physics}\ }\textbf {\bibinfo {volume}
  {145}} (\bibinfo {year} {2016}),\ 10.1063/1.4962577},\ \bibinfo {note}
  {124312},\ \Eprint
  {https://arxiv.org/abs/https://pubs.aip.org/aip/jcp/article-pdf/doi/10.1063/1.4962577/15519369/124312\_1\_online.pdf}
  {https://pubs.aip.org/aip/jcp/article-pdf/doi/10.1063/1.4962577/15519369/124312\_1\_online.pdf}
  \BibitemShut {NoStop}%
\bibitem [{\citenamefont {Dahlberg}\ \emph {et~al.}(2017)\citenamefont
  {Dahlberg}, \citenamefont {Ting}, \citenamefont {Massey}, \citenamefont
  {Allodi}, \citenamefont {Martin}, \citenamefont {Hunter},\ and\ \citenamefont
  {Engel}}]{Dahlberg2017}%
  \BibitemOpen
  \bibfield  {author} {\bibinfo {author} {\bibfnamefont {P.~D.}\ \bibnamefont
  {Dahlberg}}, \bibinfo {author} {\bibfnamefont {P.-C.}\ \bibnamefont {Ting}},
  \bibinfo {author} {\bibfnamefont {S.~C.}\ \bibnamefont {Massey}}, \bibinfo
  {author} {\bibfnamefont {M.~A.}\ \bibnamefont {Allodi}}, \bibinfo {author}
  {\bibfnamefont {E.~C.}\ \bibnamefont {Martin}}, \bibinfo {author}
  {\bibfnamefont {C.~N.}\ \bibnamefont {Hunter}},\ and\ \bibinfo {author}
  {\bibfnamefont {G.~S.}\ \bibnamefont {Engel}},\ }\bibfield  {title} {\enquote
  {\bibinfo {title} {{Mapping the ultrafast flow of harvested solar energy in
  living photosynthetic cells}},}\ }\href
  {https://doi.org/10.1038/s41467-017-01124-z} {\bibfield  {journal} {\bibinfo
  {journal} {Nature Communications}\ }\textbf {\bibinfo {volume} {8}},\
  \bibinfo {pages} {988--994} (\bibinfo {year} {2017})}\BibitemShut {NoStop}%
\bibitem [{\citenamefont {Finkelstein-Shapiro}\ \emph
  {et~al.}(2021)\citenamefont {Finkelstein-Shapiro}, \citenamefont {Mante},
  \citenamefont {Sarisozen}, \citenamefont {Wittenbecher}, \citenamefont
  {Minda}, \citenamefont {Balci}, \citenamefont {Pullerits},\ and\
  \citenamefont {Zigmantas}}]{Shapiro2021}%
  \BibitemOpen
  \bibfield  {author} {\bibinfo {author} {\bibfnamefont {D.}~\bibnamefont
  {Finkelstein-Shapiro}}, \bibinfo {author} {\bibfnamefont {P.-A.}\
  \bibnamefont {Mante}}, \bibinfo {author} {\bibfnamefont {S.}~\bibnamefont
  {Sarisozen}}, \bibinfo {author} {\bibfnamefont {L.}~\bibnamefont
  {Wittenbecher}}, \bibinfo {author} {\bibfnamefont {I.}~\bibnamefont {Minda}},
  \bibinfo {author} {\bibfnamefont {S.}~\bibnamefont {Balci}}, \bibinfo
  {author} {\bibfnamefont {T.}~\bibnamefont {Pullerits}},\ and\ \bibinfo
  {author} {\bibfnamefont {D.}~\bibnamefont {Zigmantas}},\ }\bibfield  {title}
  {\enquote {\bibinfo {title} {{Understanding radiative transitions and
  relaxation pathways in plexcitons}},}\ }\href
  {https://doi.org/https://doi.org/10.1016/j.chempr.2021.02.028} {\bibfield
  {journal} {\bibinfo  {journal} {Chem}\ }\textbf {\bibinfo {volume} {7}},\
  \bibinfo {pages} {1092--1107} (\bibinfo {year} {2021})}\BibitemShut {NoStop}%
\bibitem [{\citenamefont {Mehlenbacher}\ \emph {et~al.}(2015)\citenamefont
  {Mehlenbacher}, \citenamefont {McDonough}, \citenamefont {Grechko},
  \citenamefont {Wu}, \citenamefont {Arnold},\ and\ \citenamefont
  {Zanni}}]{Mehlenbacher2015}%
  \BibitemOpen
  \bibfield  {author} {\bibinfo {author} {\bibfnamefont {R.~D.}\ \bibnamefont
  {Mehlenbacher}}, \bibinfo {author} {\bibfnamefont {T.~J.}\ \bibnamefont
  {McDonough}}, \bibinfo {author} {\bibfnamefont {M.}~\bibnamefont {Grechko}},
  \bibinfo {author} {\bibfnamefont {M.-Y.}\ \bibnamefont {Wu}}, \bibinfo
  {author} {\bibfnamefont {M.~S.}\ \bibnamefont {Arnold}},\ and\ \bibinfo
  {author} {\bibfnamefont {M.~T.}\ \bibnamefont {Zanni}},\ }\bibfield  {title}
  {\enquote {\bibinfo {title} {{Energy transfer pathways in semiconducting
  carbon nanotubes revealed using two-dimensional white-light spectroscopy}},}\
  }\href {https://doi.org/10.1038/ncomms7732} {\bibfield  {journal} {\bibinfo
  {journal} {Nature Communications}\ }\textbf {\bibinfo {volume} {6}},\
  \bibinfo {pages} {6732} (\bibinfo {year} {2015})}\BibitemShut {NoStop}%
\bibitem [{\citenamefont {Fuller}\ and\ \citenamefont
  {Ogilvie}(2015)}]{OgilvieARPC}%
  \BibitemOpen
  \bibfield  {author} {\bibinfo {author} {\bibfnamefont {F.~D.}\ \bibnamefont
  {Fuller}}\ and\ \bibinfo {author} {\bibfnamefont {J.~P.}\ \bibnamefont
  {Ogilvie}},\ }\bibfield  {title} {\enquote {\bibinfo {title} {{Experimental
  Implementations of Two-Dimensional Fourier Transform Electronic
  Spectroscopy}},}\ }\href
  {https://doi.org/10.1146/annurev-physchem-040513-103623} {\bibfield
  {journal} {\bibinfo  {journal} {Annual Review of Physical Chemistry}\
  }\textbf {\bibinfo {volume} {66}},\ \bibinfo {pages} {667--690} (\bibinfo
  {year} {2015})}\BibitemShut {NoStop}%
\bibitem [{\citenamefont {Tiwari}(2021)}]{Tiwari2021}%
  \BibitemOpen
  \bibfield  {author} {\bibinfo {author} {\bibfnamefont {V.}~\bibnamefont
  {Tiwari}},\ }\bibfield  {title} {\enquote {\bibinfo {title}
  {{Multidimensional Electronic Spectroscopy in High-Definition - Combining
  Spectral, Temporal and Spatial Resolutions}},}\ }\href@noop {} {\bibfield
  {journal} {\bibinfo  {journal} {The Journal of Chemical Physics}\ }\textbf
  {\bibinfo {volume} {54}},\ \bibinfo {pages} {230901} (\bibinfo {year}
  {2021})}\BibitemShut {NoStop}%
\bibitem [{\citenamefont {Cirmi}\ \emph {et~al.}(2008)\citenamefont {Cirmi},
  \citenamefont {Manzoni}, \citenamefont {Brida}, \citenamefont {Silvestri},\
  and\ \citenamefont {Cerullo}}]{Cerullo2008}%
  \BibitemOpen
  \bibfield  {author} {\bibinfo {author} {\bibfnamefont {G.}~\bibnamefont
  {Cirmi}}, \bibinfo {author} {\bibfnamefont {C.}~\bibnamefont {Manzoni}},
  \bibinfo {author} {\bibfnamefont {D.}~\bibnamefont {Brida}}, \bibinfo
  {author} {\bibfnamefont {S.~D.}\ \bibnamefont {Silvestri}},\ and\ \bibinfo
  {author} {\bibfnamefont {G.}~\bibnamefont {Cerullo}},\ }\bibfield  {title}
  {\enquote {\bibinfo {title} {{Carrier-envelope phase stable,
  few-optical-cycle pulses tunable from visible to near IR}},}\ }\href
  {https://doi.org/10.1364/JOSAB.25.000B62} {\bibfield  {journal} {\bibinfo
  {journal} {J. Opt. Soc. Am. B}\ }\textbf {\bibinfo {volume} {25}},\ \bibinfo
  {pages} {B62----B69} (\bibinfo {year} {2008})}\BibitemShut {NoStop}%
\bibitem [{\citenamefont {Baltu{\v{s}}ka}, \citenamefont {Fuji},\ and\
  \citenamefont {Kobayashi}(2002)}]{Kobayashi2002}%
  \BibitemOpen
  \bibfield  {author} {\bibinfo {author} {\bibfnamefont {A.}~\bibnamefont
  {Baltu{\v{s}}ka}}, \bibinfo {author} {\bibfnamefont {T.}~\bibnamefont
  {Fuji}},\ and\ \bibinfo {author} {\bibfnamefont {T.}~\bibnamefont
  {Kobayashi}},\ }\bibfield  {title} {\enquote {\bibinfo {title} {{Visible
  pulse compression to 4 fs by optical parametric amplification and
  programmable dispersion control}},}\ }\href
  {https://doi.org/10.1364/OL.27.000306} {\bibfield  {journal} {\bibinfo
  {journal} {Opt. Lett.}\ }\textbf {\bibinfo {volume} {27}},\ \bibinfo {pages}
  {306--308} (\bibinfo {year} {2002})}\BibitemShut {NoStop}%
\bibitem [{\citenamefont {Herrmann}\ \emph {et~al.}(2010)\citenamefont
  {Herrmann}, \citenamefont {Homann}, \citenamefont {Tautz}, \citenamefont
  {Scharrer}, \citenamefont {Russell}, \citenamefont {Krausz}, \citenamefont
  {Veisz},\ and\ \citenamefont {Riedle}}]{Riedle2010}%
  \BibitemOpen
  \bibfield  {author} {\bibinfo {author} {\bibfnamefont {D.}~\bibnamefont
  {Herrmann}}, \bibinfo {author} {\bibfnamefont {C.}~\bibnamefont {Homann}},
  \bibinfo {author} {\bibfnamefont {R.}~\bibnamefont {Tautz}}, \bibinfo
  {author} {\bibfnamefont {M.}~\bibnamefont {Scharrer}}, \bibinfo {author}
  {\bibfnamefont {P.~S.}\ \bibnamefont {Russell}}, \bibinfo {author}
  {\bibfnamefont {F.}~\bibnamefont {Krausz}}, \bibinfo {author} {\bibfnamefont
  {L.}~\bibnamefont {Veisz}},\ and\ \bibinfo {author} {\bibfnamefont
  {E.}~\bibnamefont {Riedle}},\ }\bibfield  {title} {\enquote {\bibinfo {title}
  {{Approaching the full octave: Noncollinear optical parametric chirped pulse
  amplification with two-color pumping}},}\ }\href
  {https://doi.org/10.1364/OE.18.018752} {\bibfield  {journal} {\bibinfo
  {journal} {Opt. Express}\ }\textbf {\bibinfo {volume} {18}},\ \bibinfo
  {pages} {18752--18762} (\bibinfo {year} {2010})}\BibitemShut {NoStop}%
\bibitem [{\citenamefont {Tekavec}\ \emph {et~al.}(2009)\citenamefont
  {Tekavec}, \citenamefont {Myers}, \citenamefont {Lewis},\ and\ \citenamefont
  {Ogilvie}}]{Tekavec2009}%
  \BibitemOpen
  \bibfield  {author} {\bibinfo {author} {\bibfnamefont {P.~F.}\ \bibnamefont
  {Tekavec}}, \bibinfo {author} {\bibfnamefont {J.~A.}\ \bibnamefont {Myers}},
  \bibinfo {author} {\bibfnamefont {K.~L.~M.}\ \bibnamefont {Lewis}},\ and\
  \bibinfo {author} {\bibfnamefont {J.~P.}\ \bibnamefont {Ogilvie}},\
  }\bibfield  {title} {\enquote {\bibinfo {title} {{Two-dimensional electronic
  spectroscopy with a continuum probe}},}\ }\href
  {https://doi.org/10.1364/OL.34.001390} {\bibfield  {journal} {\bibinfo
  {journal} {Opt. Lett.}\ }\textbf {\bibinfo {volume} {34}},\ \bibinfo {pages}
  {1390--1392} (\bibinfo {year} {2009})}\BibitemShut {NoStop}%
\bibitem [{\citenamefont {Song}\ \emph {et~al.}(2019)\citenamefont {Song},
  \citenamefont {Konar}, \citenamefont {Sechrist}, \citenamefont {Roy},
  \citenamefont {Duan}, \citenamefont {Dziurgot}, \citenamefont {Policht},
  \citenamefont {Matutes}, \citenamefont {Kubarych},\ and\ \citenamefont
  {Ogilvie}}]{Song2019}%
  \BibitemOpen
  \bibfield  {author} {\bibinfo {author} {\bibfnamefont {Y.}~\bibnamefont
  {Song}}, \bibinfo {author} {\bibfnamefont {A.}~\bibnamefont {Konar}},
  \bibinfo {author} {\bibfnamefont {R.}~\bibnamefont {Sechrist}}, \bibinfo
  {author} {\bibfnamefont {V.~P.}\ \bibnamefont {Roy}}, \bibinfo {author}
  {\bibfnamefont {R.}~\bibnamefont {Duan}}, \bibinfo {author} {\bibfnamefont
  {J.}~\bibnamefont {Dziurgot}}, \bibinfo {author} {\bibfnamefont
  {V.}~\bibnamefont {Policht}}, \bibinfo {author} {\bibfnamefont {Y.~A.}\
  \bibnamefont {Matutes}}, \bibinfo {author} {\bibfnamefont {K.~J.}\
  \bibnamefont {Kubarych}},\ and\ \bibinfo {author} {\bibfnamefont {J.~P.}\
  \bibnamefont {Ogilvie}},\ }\bibfield  {title} {\enquote {\bibinfo {title}
  {{Multispectral multidimensional spectrometer spanning the ultraviolet to the
  mid-infrared}},}\ }\href {https://doi.org/10.1063/1.5055244} {\bibfield
  {journal} {\bibinfo  {journal} {Review of Scientific Instruments}\ }\textbf
  {\bibinfo {volume} {90}} (\bibinfo {year} {2019}),\ 10.1063/1.5055244},\
  \bibinfo {note} {013108},\ \Eprint
  {https://arxiv.org/abs/https://pubs.aip.org/aip/rsi/article-pdf/doi/10.1063/1.5055244/14754880/013108\_1\_online.pdf}
  {https://pubs.aip.org/aip/rsi/article-pdf/doi/10.1063/1.5055244/14754880/013108\_1\_online.pdf}
  \BibitemShut {NoStop}%
\bibitem [{\citenamefont {Bradler}, \citenamefont {Baum},\ and\ \citenamefont
  {Riedle}(2009)}]{Bradler2009a}%
  \BibitemOpen
  \bibfield  {author} {\bibinfo {author} {\bibfnamefont {M.}~\bibnamefont
  {Bradler}}, \bibinfo {author} {\bibfnamefont {P.}~\bibnamefont {Baum}},\ and\
  \bibinfo {author} {\bibfnamefont {E.}~\bibnamefont {Riedle}},\ }\bibfield
  {title} {\enquote {\bibinfo {title} {{Femtosecond continuum generation in
  bulk laser host materials with sub-$\mu$J pump pulses}},}\ }\href
  {https://doi.org/10.1007/s00340-009-3699-1} {\bibfield  {journal} {\bibinfo
  {journal} {Applied Physics B}\ }\textbf {\bibinfo {volume} {97}},\ \bibinfo
  {pages} {561} (\bibinfo {year} {2009})}\BibitemShut {NoStop}%
\bibitem [{\citenamefont {Bradler}\ and\ \citenamefont
  {Riedle}(2014)}]{Bradler2014}%
  \BibitemOpen
  \bibfield  {author} {\bibinfo {author} {\bibfnamefont {M.}~\bibnamefont
  {Bradler}}\ and\ \bibinfo {author} {\bibfnamefont {E.}~\bibnamefont
  {Riedle}},\ }\bibfield  {title} {\enquote {\bibinfo {title} {{Temporal and
  spectral correlations in bulk continua and improved use in transient
  spectroscopy}},}\ }\href {https://doi.org/10.1364/JOSAB.31.001465} {\bibfield
   {journal} {\bibinfo  {journal} {J. Opt. Soc. Am. B}\ }\textbf {\bibinfo
  {volume} {31}},\ \bibinfo {pages} {1465--1475} (\bibinfo {year}
  {2014})}\BibitemShut {NoStop}%
\bibitem [{\citenamefont {Son}\ \emph {et~al.}(2022)\citenamefont {Son},
  \citenamefont {Armstrong}, \citenamefont {Allen}, \citenamefont {Dhavamani},
  \citenamefont {Arnold},\ and\ \citenamefont {Zanni}}]{Son2022}%
  \BibitemOpen
  \bibfield  {author} {\bibinfo {author} {\bibfnamefont {M.}~\bibnamefont
  {Son}}, \bibinfo {author} {\bibfnamefont {Z.~T.}\ \bibnamefont {Armstrong}},
  \bibinfo {author} {\bibfnamefont {R.~T.}\ \bibnamefont {Allen}}, \bibinfo
  {author} {\bibfnamefont {A.}~\bibnamefont {Dhavamani}}, \bibinfo {author}
  {\bibfnamefont {M.~S.}\ \bibnamefont {Arnold}},\ and\ \bibinfo {author}
  {\bibfnamefont {M.~T.}\ \bibnamefont {Zanni}},\ }\bibfield  {title} {\enquote
  {\bibinfo {title} {{Energy cascades in donor-acceptor exciton-polaritons
  observed by ultrafast two-dimensional white-light spectroscopy}},}\ }\href
  {https://doi.org/10.1038/s41467-022-35046-2} {\bibfield  {journal} {\bibinfo
  {journal} {Nature Communications}\ }\textbf {\bibinfo {volume} {13}},\
  \bibinfo {pages} {7305} (\bibinfo {year} {2022})}\BibitemShut {NoStop}%
\bibitem [{\citenamefont {Kearns}\ \emph {et~al.}(2017)\citenamefont {Kearns},
  \citenamefont {Mehlenbacher}, \citenamefont {Jones},\ and\ \citenamefont
  {Zanni}}]{Kearns2017}%
  \BibitemOpen
  \bibfield  {author} {\bibinfo {author} {\bibfnamefont {N.~M.}\ \bibnamefont
  {Kearns}}, \bibinfo {author} {\bibfnamefont {R.~D.}\ \bibnamefont
  {Mehlenbacher}}, \bibinfo {author} {\bibfnamefont {A.~C.}\ \bibnamefont
  {Jones}},\ and\ \bibinfo {author} {\bibfnamefont {M.~T.}\ \bibnamefont
  {Zanni}},\ }\bibfield  {title} {\enquote {\bibinfo {title} {{Broadband 2D
  electronic spectrometer using white light and pulse shaping: noise and signal
  evaluation at 1 and 100 kHz}},}\ }\href
  {https://doi.org/10.1364/OE.25.007869} {\bibfield  {journal} {\bibinfo
  {journal} {Optics Express}\ }\textbf {\bibinfo {volume} {25}},\ \bibinfo
  {pages} {7869--7883} (\bibinfo {year} {2017})}\BibitemShut {NoStop}%
\bibitem [{\citenamefont {Kunsel}\ \emph {et~al.}(2019)\citenamefont {Kunsel},
  \citenamefont {Tiwari}, \citenamefont {Matutes}, \citenamefont {Gardiner},
  \citenamefont {Cogdell}, \citenamefont {Ogilvie},\ and\ \citenamefont
  {Jansen}}]{Kunsel2019}%
  \BibitemOpen
  \bibfield  {author} {\bibinfo {author} {\bibfnamefont {T.}~\bibnamefont
  {Kunsel}}, \bibinfo {author} {\bibfnamefont {V.}~\bibnamefont {Tiwari}},
  \bibinfo {author} {\bibfnamefont {Y.~A.}\ \bibnamefont {Matutes}}, \bibinfo
  {author} {\bibfnamefont {A.~T.}\ \bibnamefont {Gardiner}}, \bibinfo {author}
  {\bibfnamefont {R.~J.}\ \bibnamefont {Cogdell}}, \bibinfo {author}
  {\bibfnamefont {J.~P.}\ \bibnamefont {Ogilvie}},\ and\ \bibinfo {author}
  {\bibfnamefont {T.~L.~C.}\ \bibnamefont {Jansen}},\ }\bibfield  {title}
  {\enquote {\bibinfo {title} {{Simulating Fluorescence-Detected
  Two-Dimensional Electronic Spectroscopy of Multichromophoric Systems}},}\
  }\href {https://doi.org/10.1021/acs.jpcb.8b10176} {\bibfield  {journal}
  {\bibinfo  {journal} {The Journal of Physical Chemistry B}\ }\textbf
  {\bibinfo {volume} {123}},\ \bibinfo {pages} {394--406} (\bibinfo {year}
  {2019})}\BibitemShut {NoStop}%
\bibitem [{\citenamefont {Agathangelou}\ \emph {et~al.}(2021)\citenamefont
  {Agathangelou}, \citenamefont {Javed}, \citenamefont {Sessa}, \citenamefont
  {Solinas}, \citenamefont {Joffre},\ and\ \citenamefont
  {Ogilvie}}]{Ogilvie2021}%
  \BibitemOpen
  \bibfield  {author} {\bibinfo {author} {\bibfnamefont {D.}~\bibnamefont
  {Agathangelou}}, \bibinfo {author} {\bibfnamefont {A.}~\bibnamefont {Javed}},
  \bibinfo {author} {\bibfnamefont {F.}~\bibnamefont {Sessa}}, \bibinfo
  {author} {\bibfnamefont {X.}~\bibnamefont {Solinas}}, \bibinfo {author}
  {\bibfnamefont {M.}~\bibnamefont {Joffre}},\ and\ \bibinfo {author}
  {\bibfnamefont {J.~P.}\ \bibnamefont {Ogilvie}},\ }\bibfield  {title}
  {\enquote {\bibinfo {title} {{Phase-modulated rapid-scanning
  fluorescence-detected two-dimensional electronic spectroscopy}},}\ }\href
  {https://doi.org/10.1063/5.0057649} {\bibfield  {journal} {\bibinfo
  {journal} {The Journal of Chemical Physics}\ }\textbf {\bibinfo {volume}
  {155}} (\bibinfo {year} {2021}),\ 10.1063/5.0057649},\ \bibinfo {note}
  {094201},\ \Eprint
  {https://arxiv.org/abs/https://pubs.aip.org/aip/jcp/article-pdf/doi/10.1063/5.0057649/15630235/094201\_1\_online.pdf}
  {https://pubs.aip.org/aip/jcp/article-pdf/doi/10.1063/5.0057649/15630235/094201\_1\_online.pdf}
  \BibitemShut {NoStop}%
\bibitem [{\citenamefont {Sahu}\ \emph {et~al.}(2023)\citenamefont {Sahu},
  \citenamefont {Bhat}, \citenamefont {Patra},\ and\ \citenamefont
  {Tiwari}}]{Sahu2023}%
  \BibitemOpen
  \bibfield  {author} {\bibinfo {author} {\bibfnamefont {A.}~\bibnamefont
  {Sahu}}, \bibinfo {author} {\bibfnamefont {V.~N.}\ \bibnamefont {Bhat}},
  \bibinfo {author} {\bibfnamefont {S.}~\bibnamefont {Patra}},\ and\ \bibinfo
  {author} {\bibfnamefont {V.}~\bibnamefont {Tiwari}},\ }\bibfield  {title}
  {\enquote {\bibinfo {title} {{High-sensitivity fluorescence-detected
  multidimensional electronic spectroscopy through continuous pump–probe
  delay scan}},}\ }\href {https://doi.org/10.1063/5.0130887} {\bibfield
  {journal} {\bibinfo  {journal} {The Journal of Chemical Physics}\ }\textbf
  {\bibinfo {volume} {158}},\ \bibinfo {pages} {024201} (\bibinfo {year}
  {2023})},\ \Eprint
  {https://arxiv.org/abs/https://pubs.aip.org/aip/jcp/article-pdf/doi/10.1063/5.0130887/16668728/024201\_1\_online.pdf}
  {https://pubs.aip.org/aip/jcp/article-pdf/doi/10.1063/5.0130887/16668728/024201\_1\_online.pdf}
  \BibitemShut {NoStop}%
\bibitem [{\citenamefont {Zheng}\ \emph {et~al.}(2014)\citenamefont {Zheng},
  \citenamefont {Caram}, \citenamefont {Dahlberg}, \citenamefont {Rolczynski},
  \citenamefont {Viswanathan}, \citenamefont {Dolzhnikov}, \citenamefont
  {Khadivi}, \citenamefont {Talapin},\ and\ \citenamefont {Engel}}]{Engel2014}%
  \BibitemOpen
  \bibfield  {author} {\bibinfo {author} {\bibfnamefont {H.}~\bibnamefont
  {Zheng}}, \bibinfo {author} {\bibfnamefont {J.~R.}\ \bibnamefont {Caram}},
  \bibinfo {author} {\bibfnamefont {P.~D.}\ \bibnamefont {Dahlberg}}, \bibinfo
  {author} {\bibfnamefont {B.~S.}\ \bibnamefont {Rolczynski}}, \bibinfo
  {author} {\bibfnamefont {S.}~\bibnamefont {Viswanathan}}, \bibinfo {author}
  {\bibfnamefont {D.~S.}\ \bibnamefont {Dolzhnikov}}, \bibinfo {author}
  {\bibfnamefont {A.}~\bibnamefont {Khadivi}}, \bibinfo {author} {\bibfnamefont
  {D.~V.}\ \bibnamefont {Talapin}},\ and\ \bibinfo {author} {\bibfnamefont
  {G.~S.}\ \bibnamefont {Engel}},\ }\bibfield  {title} {\enquote {\bibinfo
  {title} {{Dispersion-free continuum two-dimensional electronic
  spectrometer}},}\ }\href {https://doi.org/10.1364/AO.53.001909} {\bibfield
  {journal} {\bibinfo  {journal} {Appl. Opt.}\ }\textbf {\bibinfo {volume}
  {53}},\ \bibinfo {pages} {1909--1917} (\bibinfo {year} {2014})}\BibitemShut
  {NoStop}%
\bibitem [{\citenamefont {Spokoyny}, \citenamefont {Koh},\ and\ \citenamefont
  {Harel}(2015)}]{Harel2015}%
  \BibitemOpen
  \bibfield  {author} {\bibinfo {author} {\bibfnamefont {B.}~\bibnamefont
  {Spokoyny}}, \bibinfo {author} {\bibfnamefont {C.~J.}\ \bibnamefont {Koh}},\
  and\ \bibinfo {author} {\bibfnamefont {E.}~\bibnamefont {Harel}},\ }\bibfield
   {title} {\enquote {\bibinfo {title} {{Stable and high-power few cycle
  supercontinuum for 2D ultrabroadband electronic spectroscopy}},}\ }\href
  {https://doi.org/10.1364/OL.40.001014} {\bibfield  {journal} {\bibinfo
  {journal} {Opt. Lett.}\ }\textbf {\bibinfo {volume} {40}},\ \bibinfo {pages}
  {1014--1017} (\bibinfo {year} {2015})}\BibitemShut {NoStop}%
\bibitem [{\citenamefont {Son}, \citenamefont {Mosquera-V{\'{a}}zquez},\ and\
  \citenamefont {Schlau-Cohen}(2017)}]{Cohen2017}%
  \BibitemOpen
  \bibfield  {author} {\bibinfo {author} {\bibfnamefont {M.}~\bibnamefont
  {Son}}, \bibinfo {author} {\bibfnamefont {S.}~\bibnamefont
  {Mosquera-V{\'{a}}zquez}},\ and\ \bibinfo {author} {\bibfnamefont {G.~S.}\
  \bibnamefont {Schlau-Cohen}},\ }\bibfield  {title} {\enquote {\bibinfo
  {title} {{Ultrabroadband 2D electronic spectroscopy with high-speed,
  shot-to-shot detection}},}\ }\href {https://doi.org/10.1364/OE.25.018950}
  {\bibfield  {journal} {\bibinfo  {journal} {Opt. Express}\ }\textbf {\bibinfo
  {volume} {25}},\ \bibinfo {pages} {18950--18962} (\bibinfo {year}
  {2017})}\BibitemShut {NoStop}%
\bibitem [{\citenamefont {Kanal}\ \emph {et~al.}(2014)\citenamefont {Kanal},
  \citenamefont {Keiber}, \citenamefont {Eck},\ and\ \citenamefont
  {Brixner}}]{Brixner2014}%
  \BibitemOpen
  \bibfield  {author} {\bibinfo {author} {\bibfnamefont {F.}~\bibnamefont
  {Kanal}}, \bibinfo {author} {\bibfnamefont {S.}~\bibnamefont {Keiber}},
  \bibinfo {author} {\bibfnamefont {R.}~\bibnamefont {Eck}},\ and\ \bibinfo
  {author} {\bibfnamefont {T.}~\bibnamefont {Brixner}},\ }\bibfield  {title}
  {\enquote {\bibinfo {title} {{100-kHz shot-to-shot broadband data acquisition
  for high-repetition-rate pump--probe spectroscopy}},}\ }\href
  {https://doi.org/10.1364/OE.22.016965} {\bibfield  {journal} {\bibinfo
  {journal} {Opt. Express}\ }\textbf {\bibinfo {volume} {22}},\ \bibinfo
  {pages} {16965--16975} (\bibinfo {year} {2014})}\BibitemShut {NoStop}%
\bibitem [{\citenamefont {Bhat}\ \emph {et~al.}(2023)\citenamefont {Bhat},
  \citenamefont {Thomas}, \citenamefont {Bhattacharyya},\ and\ \citenamefont
  {Tiwari}}]{Bhat2023}%
  \BibitemOpen
  \bibfield  {author} {\bibinfo {author} {\bibfnamefont {V.~N.}\ \bibnamefont
  {Bhat}}, \bibinfo {author} {\bibfnamefont {A.~S.}\ \bibnamefont {Thomas}},
  \bibinfo {author} {\bibfnamefont {A.}~\bibnamefont {Bhattacharyya}},\ and\
  \bibinfo {author} {\bibfnamefont {V.}~\bibnamefont {Tiwari}},\ }\bibfield
  {title} {\enquote {\bibinfo {title} {Rapid scan white light
  pump\&\#x2013;probe spectroscopy with 100 khz shot-to-shot detection},}\
  }\href {https://doi.org/10.1364/OPTCON.496928} {\bibfield  {journal}
  {\bibinfo  {journal} {Opt. Continuum}\ }\textbf {\bibinfo {volume} {2}},\
  \bibinfo {pages} {1981--1995} (\bibinfo {year} {2023})}\BibitemShut {NoStop}%
\bibitem [{\citenamefont {Moon}(1993)}]{Moon1993}%
  \BibitemOpen
  \bibfield  {author} {\bibinfo {author} {\bibfnamefont {J.~A.}\ \bibnamefont
  {Moon}},\ }\bibfield  {title} {\enquote {\bibinfo {title} {{Optimization of
  signal‐to‐noise ratios in pump‐probe spectroscopy}},}\ }\href
  {https://doi.org/10.1063/1.1144009} {\bibfield  {journal} {\bibinfo
  {journal} {Review of Scientific Instruments}\ }\textbf {\bibinfo {volume}
  {64}},\ \bibinfo {pages} {1775--1778} (\bibinfo {year} {1993})}\BibitemShut
  {NoStop}%
\bibitem [{\citenamefont {Tull}, \citenamefont {Dugan},\ and\ \citenamefont
  {Warren}(1997)}]{Tull1997}%
  \BibitemOpen
  \bibfield  {author} {\bibinfo {author} {\bibfnamefont {J.~X.}\ \bibnamefont
  {Tull}}, \bibinfo {author} {\bibfnamefont {M.~A.}\ \bibnamefont {Dugan}},\
  and\ \bibinfo {author} {\bibfnamefont {W.~S.}\ \bibnamefont {Warren}},\
  }\enquote {\bibinfo {title} {{High-resolution, ultrafast laser pulse shaping
  and its applications}},}\ in\ \href
  {https://doi.org/https://doi.org/10.1016/S1057-2732(97)80002-2} {\emph
  {\bibinfo {booktitle} {Advances in Magnetic and Optical Resonance}}},\
  Vol.~\bibinfo {volume} {20},\ \bibinfo {editor} {edited by\ \bibinfo {editor}
  {\bibfnamefont {W.~S.}\ \bibnamefont {Warren}}}\ (\bibinfo  {publisher}
  {Academic Press},\ \bibinfo {year} {1997})\ pp.\ \bibinfo {pages}
  {1--II}\BibitemShut {NoStop}%
\bibitem [{\citenamefont {Weiner}(2011)}]{Weiner2011}%
  \BibitemOpen
  \bibfield  {author} {\bibinfo {author} {\bibfnamefont {A.~M.}\ \bibnamefont
  {Weiner}},\ }\bibfield  {title} {\enquote {\bibinfo {title} {{Ultrafast
  optical pulse shaping: A tutorial review}},}\ }\href
  {https://doi.org/https://doi.org/10.1016/j.optcom.2011.03.084} {\bibfield
  {journal} {\bibinfo  {journal} {Optics Communications}\ }\textbf {\bibinfo
  {volume} {284}},\ \bibinfo {pages} {3669--3692} (\bibinfo {year}
  {2011})}\BibitemShut {NoStop}%
\bibitem [{\citenamefont {Luther}\ \emph {et~al.}(2016)\citenamefont {Luther},
  \citenamefont {Tracy}, \citenamefont {Gerrity}, \citenamefont {Brown},\ and\
  \citenamefont {Krummel}}]{Krummel2016}%
  \BibitemOpen
  \bibfield  {author} {\bibinfo {author} {\bibfnamefont {B.~M.}\ \bibnamefont
  {Luther}}, \bibinfo {author} {\bibfnamefont {K.~M.}\ \bibnamefont {Tracy}},
  \bibinfo {author} {\bibfnamefont {M.}~\bibnamefont {Gerrity}}, \bibinfo
  {author} {\bibfnamefont {S.}~\bibnamefont {Brown}},\ and\ \bibinfo {author}
  {\bibfnamefont {A.~T.}\ \bibnamefont {Krummel}},\ }\bibfield  {title}
  {\enquote {\bibinfo {title} {{2D IR spectroscopy at 100 kHz utilizing a
  Mid-IR OPCPA laser source}},}\ }\href {https://doi.org/10.1364/OE.24.004117}
  {\bibfield  {journal} {\bibinfo  {journal} {Opt. Express}\ }\textbf {\bibinfo
  {volume} {24}},\ \bibinfo {pages} {4117--4127} (\bibinfo {year}
  {2016})}\BibitemShut {NoStop}%
\bibitem [{\citenamefont {Jones}\ \emph {et~al.}(2020)\citenamefont {Jones},
  \citenamefont {Kearns}, \citenamefont {Ho}, \citenamefont {Flach},\ and\
  \citenamefont {Zanni}}]{Jones2020}%
  \BibitemOpen
  \bibfield  {author} {\bibinfo {author} {\bibfnamefont {A.~C.}\ \bibnamefont
  {Jones}}, \bibinfo {author} {\bibfnamefont {N.~M.}\ \bibnamefont {Kearns}},
  \bibinfo {author} {\bibfnamefont {J.-J.}\ \bibnamefont {Ho}}, \bibinfo
  {author} {\bibfnamefont {J.~T.}\ \bibnamefont {Flach}},\ and\ \bibinfo
  {author} {\bibfnamefont {M.~T.}\ \bibnamefont {Zanni}},\ }\bibfield  {title}
  {\enquote {\bibinfo {title} {{Impact of non-equilibrium molecular packings on
  singlet fission in microcrystals observed using 2D white-light
  microscopy}},}\ }\href {https://doi.org/10.1038/s41557-019-0368-9} {\bibfield
   {journal} {\bibinfo  {journal} {Nature Chemistry}\ }\textbf {\bibinfo
  {volume} {12}},\ \bibinfo {pages} {40--47} (\bibinfo {year}
  {2020})}\BibitemShut {NoStop}%
\bibitem [{\citenamefont {Martin}\ \emph {et~al.}()\citenamefont {Martin},
  \citenamefont {Horng}, \citenamefont {Ruth}, \citenamefont {Paik},
  \citenamefont {Wentzel}, \citenamefont {Deng},\ and\ \citenamefont
  {Cundiff}}]{Martin2020}%
  \BibitemOpen
  \bibfield  {author} {\bibinfo {author} {\bibfnamefont {E.~W.}\ \bibnamefont
  {Martin}}, \bibinfo {author} {\bibfnamefont {J.}~\bibnamefont {Horng}},
  \bibinfo {author} {\bibfnamefont {H.~G.}\ \bibnamefont {Ruth}}, \bibinfo
  {author} {\bibfnamefont {E.}~\bibnamefont {Paik}}, \bibinfo {author}
  {\bibfnamefont {M.-H.}\ \bibnamefont {Wentzel}}, \bibinfo {author}
  {\bibfnamefont {H.}~\bibnamefont {Deng}},\ and\ \bibinfo {author}
  {\bibfnamefont {S.~T.}\ \bibnamefont {Cundiff}},\ }\bibfield  {title}
  {\enquote {\bibinfo {title} {{Encapsulation Narrows and Preserves the
  Excitonic Homogeneous Linewidth of Exfoliated Monolayer MoSe2}, url =
  {https://link.aps.org/doi/10.1103/PhysRevApplied.14.021002}, volume = {14},
  year = {2020}},}\ }\href {https://doi.org/10.1103/PhysRevApplied.14.021002}
  {\bibinfo  {journal} {Phys. Rev. Applied}\ ,\ \bibinfo {pages}
  {21002}}\BibitemShut {NoStop}%
\bibitem [{\citenamefont {Goetz}\ \emph {et~al.}(2018)\citenamefont {Goetz},
  \citenamefont {Li}, \citenamefont {Kolb}, \citenamefont {Pflaum},\ and\
  \citenamefont {Brixner}}]{Goetz2018}%
  \BibitemOpen
\bibfield  {journal} {  }\bibfield  {author} {\bibinfo {author} {\bibfnamefont
  {S.}~\bibnamefont {Goetz}}, \bibinfo {author} {\bibfnamefont
  {D.}~\bibnamefont {Li}}, \bibinfo {author} {\bibfnamefont {V.}~\bibnamefont
  {Kolb}}, \bibinfo {author} {\bibfnamefont {J.}~\bibnamefont {Pflaum}},\ and\
  \bibinfo {author} {\bibfnamefont {T.}~\bibnamefont {Brixner}},\ }\bibfield
  {title} {\enquote {\bibinfo {title} {{Coherent two-dimensional fluorescence
  micro-spectroscopy}},}\ }\href {https://doi.org/10.1364/OE.26.003915}
  {\bibfield  {journal} {\bibinfo  {journal} {Optics Express}\ }\textbf
  {\bibinfo {volume} {26}},\ \bibinfo {pages} {3915--3925} (\bibinfo {year}
  {2018})}\BibitemShut {NoStop}%
\bibitem [{\citenamefont {Tiwari}\ \emph {et~al.}(2018)\citenamefont {Tiwari},
  \citenamefont {Matutes}, \citenamefont {Gardiner}, \citenamefont {Jansen},
  \citenamefont {Cogdell},\ and\ \citenamefont {Ogilvie}}]{Tiwari2018a}%
  \BibitemOpen
  \bibfield  {author} {\bibinfo {author} {\bibfnamefont {V.}~\bibnamefont
  {Tiwari}}, \bibinfo {author} {\bibfnamefont {Y.~A.}\ \bibnamefont {Matutes}},
  \bibinfo {author} {\bibfnamefont {A.~T.}\ \bibnamefont {Gardiner}}, \bibinfo
  {author} {\bibfnamefont {T.~L.~C.}\ \bibnamefont {Jansen}}, \bibinfo {author}
  {\bibfnamefont {R.~J.}\ \bibnamefont {Cogdell}},\ and\ \bibinfo {author}
  {\bibfnamefont {J.~P.}\ \bibnamefont {Ogilvie}},\ }\bibfield  {title}
  {\enquote {\bibinfo {title} {{Spatially-resolved fluorescence-detected
  two-dimensional electronic spectroscopy probes varying excitonic structure in
  photosynthetic bacteria}},}\ }\href
  {https://doi.org/10.1038/s41467-018-06619-x} {\bibfield  {journal} {\bibinfo
  {journal} {Nature Communications}\ }\textbf {\bibinfo {volume} {9}},\
  \bibinfo {pages} {4219} (\bibinfo {year} {2018})}\BibitemShut {NoStop}%
\bibitem [{\citenamefont {R{\'{e}}hault}\ \emph {et~al.}(2014)\citenamefont
  {R{\'{e}}hault}, \citenamefont {Maiuri}, \citenamefont {Oriana},\ and\
  \citenamefont {Cerullo}}]{Cerullo2014}%
  \BibitemOpen
  \bibfield  {author} {\bibinfo {author} {\bibfnamefont {J.}~\bibnamefont
  {R{\'{e}}hault}}, \bibinfo {author} {\bibfnamefont {M.}~\bibnamefont
  {Maiuri}}, \bibinfo {author} {\bibfnamefont {A.}~\bibnamefont {Oriana}},\
  and\ \bibinfo {author} {\bibfnamefont {G.}~\bibnamefont {Cerullo}},\
  }\bibfield  {title} {\enquote {\bibinfo {title} {{Two-dimensional electronic
  spectroscopy with birefringent wedges}},}\ }\href
  {https://doi.org/10.1063/1.4902938} {\bibfield  {journal} {\bibinfo
  {journal} {Review of Scientific Instruments}\ }\textbf {\bibinfo {volume}
  {85}},\ \bibinfo {pages} {123107} (\bibinfo {year} {2014})}\BibitemShut
  {NoStop}%
\bibitem [{\citenamefont {Donley}\ \emph {et~al.}(2005)\citenamefont {Donley},
  \citenamefont {Heavner}, \citenamefont {Levi}, \citenamefont {Tataw},\ and\
  \citenamefont {Jefferts}}]{Jefferts2005}%
  \BibitemOpen
  \bibfield  {author} {\bibinfo {author} {\bibfnamefont {E.~A.}\ \bibnamefont
  {Donley}}, \bibinfo {author} {\bibfnamefont {T.~P.}\ \bibnamefont {Heavner}},
  \bibinfo {author} {\bibfnamefont {F.}~\bibnamefont {Levi}}, \bibinfo {author}
  {\bibfnamefont {M.~O.}\ \bibnamefont {Tataw}},\ and\ \bibinfo {author}
  {\bibfnamefont {S.~R.}\ \bibnamefont {Jefferts}},\ }\bibfield  {title}
  {\enquote {\bibinfo {title} {{Double-pass acousto-optic modulator system}},}\
  }\href {https://doi.org/10.1063/1.1930095} {\bibfield  {journal} {\bibinfo
  {journal} {Review of Scientific Instruments}\ }\textbf {\bibinfo {volume}
  {76}},\ \bibinfo {pages} {63112} (\bibinfo {year} {2005})}\BibitemShut
  {NoStop}%
\bibitem [{\citenamefont {Cho}\ \emph {et~al.}(2013)\citenamefont {Cho},
  \citenamefont {Tiwari}, \citenamefont {Hill}, \citenamefont {Peters},
  \citenamefont {Courtney}, \citenamefont {Spencer},\ and\ \citenamefont
  {Jonas}}]{Cho2013}%
  \BibitemOpen
  \bibfield  {author} {\bibinfo {author} {\bibfnamefont {B.}~\bibnamefont
  {Cho}}, \bibinfo {author} {\bibfnamefont {V.}~\bibnamefont {Tiwari}},
  \bibinfo {author} {\bibfnamefont {R.~J.}\ \bibnamefont {Hill}}, \bibinfo
  {author} {\bibfnamefont {W.~K.}\ \bibnamefont {Peters}}, \bibinfo {author}
  {\bibfnamefont {T.~L.}\ \bibnamefont {Courtney}}, \bibinfo {author}
  {\bibfnamefont {A.~P.}\ \bibnamefont {Spencer}},\ and\ \bibinfo {author}
  {\bibfnamefont {D.~M.}\ \bibnamefont {Jonas}},\ }\bibfield  {title} {\enquote
  {\bibinfo {title} {{Absolute Measurement of Femtosecond Pump–Probe Signal
  Strength}},}\ }\href {https://doi.org/10.1021/jp4019662} {\bibfield
  {journal} {\bibinfo  {journal} {The Journal of Physical Chemistry A}\
  }\textbf {\bibinfo {volume} {117}},\ \bibinfo {pages} {6332--6345} (\bibinfo
  {year} {2013})}\BibitemShut {NoStop}%
\bibitem [{\citenamefont {Gallagher~Faeder}\ and\ \citenamefont
  {Jonas}(1999)}]{Jonas1999}%
  \BibitemOpen
  \bibfield  {author} {\bibinfo {author} {\bibfnamefont {S.~M.}\ \bibnamefont
  {Gallagher~Faeder}}\ and\ \bibinfo {author} {\bibfnamefont {D.~M.}\
  \bibnamefont {Jonas}},\ }\bibfield  {title} {\enquote {\bibinfo {title}
  {Two-dimensional electronic correlation and relaxation spectra: Theory and
  model calculations},}\ }\href {https://doi.org/10.1021/jp9925738} {\bibfield
  {journal} {\bibinfo  {journal} {The Journal of Physical Chemistry A}\
  }\textbf {\bibinfo {volume} {103}},\ \bibinfo {pages} {10489--10505}
  (\bibinfo {year} {1999})},\ \Eprint
  {https://arxiv.org/abs/https://doi.org/10.1021/jp9925738}
  {https://doi.org/10.1021/jp9925738} \BibitemShut {NoStop}%
\bibitem [{\citenamefont {Piland}\ and\ \citenamefont
  {Grumstrup}(2019)}]{Grumstrup2019}%
  \BibitemOpen
  \bibfield  {author} {\bibinfo {author} {\bibfnamefont {G.}~\bibnamefont
  {Piland}}\ and\ \bibinfo {author} {\bibfnamefont {E.~M.}\ \bibnamefont
  {Grumstrup}},\ }\bibfield  {title} {\enquote {\bibinfo {title}
  {High-repetition rate broadband pump–probe microscopy},}\ }\href
  {https://doi.org/10.1021/acs.jpca.9b03858} {\bibfield  {journal} {\bibinfo
  {journal} {The Journal of Physical Chemistry A}\ }\textbf {\bibinfo {volume}
  {123}},\ \bibinfo {pages} {8709--8716} (\bibinfo {year} {2019})},\ \bibinfo
  {note} {pMID: 31251048},\ \Eprint
  {https://arxiv.org/abs/https://doi.org/10.1021/acs.jpca.9b03858}
  {https://doi.org/10.1021/acs.jpca.9b03858} \BibitemShut {NoStop}%
\bibitem [{ima()}]{imaq}%
  \BibitemOpen
  \bibfield  {title} {\enquote {\bibinfo {title} {{National Instruments
  IMAQdx}},}\ }\href
  {https://www.ni.com/docs/en-US/bundle/ni-imaqdx-vi-ref/page/ni-imaqdx{\_}vi{\_}reference/ni-imaqdx{\_}pal.html}
  {\bibinfo  {journal}
  {https://www.ni.com/docs/en-US/bundle/ni-imaqdx-vi-ref/page/ni-imaqdx{\_}vi{\_}reference/ni-imaqdx{\_}pal.html}\
  }\BibitemShut {NoStop}%
\bibitem [{\citenamefont {Hybl}, \citenamefont {Albrecht~Ferro},\ and\
  \citenamefont {Jonas}(2001)}]{Jonas2001}%
  \BibitemOpen
\bibfield  {journal} {  }\bibfield  {author} {\bibinfo {author} {\bibfnamefont
  {J.~D.}\ \bibnamefont {Hybl}}, \bibinfo {author} {\bibfnamefont
  {A.}~\bibnamefont {Albrecht~Ferro}},\ and\ \bibinfo {author} {\bibfnamefont
  {D.~M.}\ \bibnamefont {Jonas}},\ }\bibfield  {title} {\enquote {\bibinfo
  {title} {{Two-dimensional Fourier transform electronic spectroscopy}},}\
  }\href {https://doi.org/10.1063/1.1398579} {\bibfield  {journal} {\bibinfo
  {journal} {The Journal of Chemical Physics}\ }\textbf {\bibinfo {volume}
  {115}},\ \bibinfo {pages} {6606--6622} (\bibinfo {year} {2001})},\ \Eprint
  {https://arxiv.org/abs/https://pubs.aip.org/aip/jcp/article-pdf/115/14/6606/10835542/6606\_1\_online.pdf}
  {https://pubs.aip.org/aip/jcp/article-pdf/115/14/6606/10835542/6606\_1\_online.pdf}
  \BibitemShut {NoStop}%
\end{thebibliography}%

\end{document}